\documentclass[lettersize,journal]{IEEEtran}
\usepackage{color}
\usepackage{amsfonts,amsmath,amssymb,amsthm}
\usepackage{latexsym,amscd}
\usepackage{amsbsy}
\usepackage{graphicx,multirow,bm}
\usepackage{algorithm}
\usepackage{algorithmicx}
\usepackage{algpseudocode}
\usepackage{xcolor}
\usepackage{tikz}
\usepackage{diagbox}
\usepackage{multicol}
\usepackage{arydshln}
\usepackage{booktabs}
\usepackage{bbm}
\usepackage{caption}
\usepackage{times}
\usepackage{amsmath}
\usepackage{cases}
\usepackage{subcaption}
\usepackage{cite}
\usepackage{hyperref}
\usepackage{array}

\allowdisplaybreaks[4]
\makeatletter
\def\@cite#1#2{\text{[{#1\if@tempswa, #2\fi}]}} 

\newcommand{\mb}{\mathbf}

\newtheorem{theorem}{Theorem}
\newtheorem{lemma}{Lemma}

\newtheorem{corollary}{Corollary}
\newtheorem{definition}{Definition}
\newtheorem{example}{Example}
\newtheorem{remark}{Remark}

\newtheorem{construction}{Construction}

\newtheorem{observation}{Observation}
\begin{document}

	\title{Error-Correcting Codes for Two Bursts of $t_1$-Deletion-$t_2$-Insertion with Reduced Complexity}
	\author{Yajuan Liu,~\IEEEmembership{Member,~IEEE}, and Tolga M. Duman, \IEEEmembership{Fellow, IEEE}
		\thanks{
			Y. Liu and T. M. Duman are with the Electrical and Electronics Engineering Department, Bilkent University, Ankara, Turkey, 06800 (email: yajuan.liu@bilkent.edu.tr, duman@ee.bilkent.edu.tr).}
		
		\thanks{Part of this work was presented at the 2026 IEEE International Symposium on Information Theory (ISIT 2026)  \cite{liu2026two}.}
	}

	\maketitle
	
	\begin{abstract}
		Burst errors involving simultaneous insertions, deletions, and substitutions occur in practical scenarios, including DNA data storage and document synchronization, motivating the development of channel codes that can correct such errors. In this paper, we construct error-correcting codes (ECCs) capable of handling multiple bursts of $t_1$-deletion-$t_2$-insertion ($(t_1,t_2)$-DI) errors, where each burst consists of $t_1$ deletions followed by $t_2$ insertions in a binary sequence. 
		We make three key contributions: First, we establish the fundamental equivalence among (i) ECCs correcting two bursts of $(t_1,t_2)$-DI errors, (ii) ECCs correcting  two bursts of $(t_2,t_1)$-DI errors, and (iii) ECCs correcting one burst of $(t_1,t_2)$-DI together with one burst of $(t_2,t_1)$-DI errors. Then, we derive lower and upper bounds on the code size of two-burst $(t_1,t_2)$-DI ECCs, which can naturally be extended to the case of multiple bursts. Finally,  we present constructions of ECCs correcting two bursts of $(t_1,t_2)$-DI errors. 
		Compared with codes obtained via the direct application of the syndrome compression technique, the proposed constructions achieve substantially improved computational efficiency.
	\end{abstract}
	
	\section{Introduction}
	
	\IEEEPARstart{C}{hannels} with insertions and deletions, referred to as synchronization error channels, model certain communication systems, including magnetic recording channels \cite{levenshtein1993perfect}, document synchronization applications \cite{cheng2019block}, and DNA data storage systems \cite{heckel2019acharacterization}. 
	Over the past half-century, there have been many developments addressing different aspects of such systems. For instance, different types of channel codes have been developed for correcting deletion and insertion errors \cite{levenshtein1965binary,bours1994construction,hu2010achievable,paluncic2012multiple}.
	Although there is extensive work on basic deletion/insertion models, there are many other scenarios that have not been addressed. 
	Of particular interest is the presence of insertion, deletion, and substitution errors occurring together. 
	With this motivation, there have been recent works on developing error-correcting codes (ECCs) capable of effectively addressing more general channels with synchronization errors.
	
	Despite many recent developments, the design of ECCs for synchronization errors remains a fundamental challenge. 
	Going back to the foundations, the seminal work by Levenshtein \cite{levenshtein1965binary} constructed a single-edit (insertion/deletion/substitution) correcting code requiring only $\log n + 1$ bits of redundancy, using the well-known Varshamov-Tenengolts (VT) syndrome \cite{varshamov1965code}.\footnote{Unless stated otherwise, all logarithmic operations in this paper are base $2$. For simplicity, we omit the ceiling operator throughout the paper.} 
	While there were only sporadic results for a long time, recent breakthroughs by Brakensiek \textit{et al.} \cite{brakensiek2018efficient} on $t$-deletion codes have spurred substantial progress in this field over the last decade \cite{gabrys2019codes,sima2020two,guruswami2021explicit,sima2020optimal,sima2021on,sima2020optimalcodes,pi2025two,sun2024binary,cai2021correcting,song2023nonbinary,liu2024explicit,song2022systematic,schaller2025a}.
	For the specific case of two deletions,  the state-of-the-art result was developed by Guruswami \textit{et al.} \cite{guruswami2021explicit} for binary codes with $4\log n + o(\log n)$ bits of redundancy. 
	This result was extended to non-binary alphabets by Song \textit{et al.} \cite{song2023nonbinary} and Liu \textit{et al.} \cite{liu2024explicit}, who introduced codes with $5\log n + o(\log n)$ bits of redundancy under certain parameter settings.
	For the general case of $t$-deletions with $t\ge 2$,  Sima \textit{et al.} \cite{sima2021on,sima2020optimal} developed binary codes with $4t\log n + o(\log n)$ bits of redundancy using the syndrome compression technique \cite{sima2020syndrome}. 
	Subsequent work by Song \textit{et al.} \cite{song2022systematic} improved this result to $(4t-1)\log n + o(\log n)$ through pre-coding methods. 
	Regarding the $q$-ary case, Sima  \textit{et al.} \cite{sima2020optimalcodes} initially achieved a redundancy of at most $30t\log n + o(\log n)$ bits for $q$-ary $t$-deletion codes, and Schaller \textit{et al.} \cite{schaller2025a} recently improved this result to $t\log q + (3t-1)\log n + o(\log n)$ bits of redundancy for sufficiently large $n$ and $q > n^{2+\epsilon}$, where $\epsilon > 0$.
	
	In some practical scenarios, such as  file synchronization \cite{venkataramanan2015low} and wireless sensor networks \cite{jeong2003forward}, data corruption may occur as bursts of insertions, deletions, and substitutions. Motivated by this, significant research has focused on designing ECCs capable of handling such errors as well. Several works have investigated code constructions for correcting a single burst of $t$-deletions or up to $t$-deletions, including \cite{schoeny2017codes,lenz2020optimal,song2023nonbinary,song2024new,wang2024nonbinary}. Among them, the recent coding scheme proposed by Song \textit{et al.} \cite{song2024new} achieves a redundancy of $\log n + 8\log\log n+o(\log \log n)$ bits for correcting a single burst of at most $t$-deletions.  
	As a further extension in this direction, Sun \textit{et al.} \cite{sun2025codes} proposed ECCs with $\log n + o(\log n)$ bits of redundancy to correct a single burst of edits in $q$-ary sequences ($q \geq 2$). 
	Beyond the single-burst setting, Sima \textit{et al.} \cite{sima2020syndrome} constructed ECCs capable of correcting $m$ bursts of at most $t$-deletions with a redundancy of $4m\log n+o(\log n)$ bits by employing the syndrome compression technique. 
	Subsequently, focusing on the case of two bursts of exactly $t$-deletions, Ye \textit{et al.} \cite{ye2024codes} presented explicit ECC constructions with $5\log n+o(\log n)$ bits of redundancy.

	In real-world applications, multiple error types may occur simultaneously.
	Early work by Smagloy \textit{et al.} \cite{smagloy2020singleseletion} addressed this challenge by constructing binary ECCs correcting a single deletion and a single substitution. 
	Later, Song \textit{et al.} \cite{song2022systematic} generalized this result to multiple deletions and multiple substitutions under both binary and non-binary settings.  
	A notable contribution by Schoeny \textit{et al.} \cite{schoeny2017codes} introduced burst ECCs capable of handling both deletions and insertions simultaneously, specifically constructing binary codes for one burst of $2$-deletions-$1$-insertion (referred to as $(2,1)$-DI) errors. 
	This result was later extended by Lu \textit{et al.} to one-burst $(t_1,1)$-DI ECCs \cite{lu2022tdeletion1} and subsequently to $(t_1,t_2)$-DI ECCs with $t_1\ge 2t_2$ \cite{lu2023tdeletion}, thereby establishing a lower bound of $\log(n-t_1+2)+t_1-1$ bits of redundancy for such codes.  
	Most recently, Sun \textit{et al.} \cite{sun2024asymptotically} provided explicit constructions of $q$-ary one-burst $(t_1,t_2)$-DI ECCs with $\log n + O(1)$ bits of redundancy, applicable for any nonnegative integers $t_1,t_2$ and $q \geq 2$. 

		Despite the developments for correcting different types of synchronization errors, the study of channels involving multiple simultaneous burst errors remains largely open and technically challenging. 
		To the best of our knowledge, prior results addressing two bursts are restricted to deletion-only models. 
		Our work extends this line of research to a substantially more general setting that simultaneously considers two bursts of $(t_1,t_2)$-DI.
		In particular,	when $t_1=0$, the proposed $(t_1,t_2)$-DI model reduces to two bursts of insertions; 
		when $t_2=0$, it reduces to two bursts of deletions; and when $t_1=t_2$, it corresponds to two bursts of substitutions.
	
		Under the generalized model, we first establish the equivalence among the following classes of ECCs: codes correcting two bursts of $(t_1,t_2)$-DI errors, codes correcting two bursts of $(t_2,t_1)$-DI errors, and codes correcting one burst of $(t_1,t_2)$-DI together with one burst of $(t_2,t_1)$-DI errors. 
	We then derive both lower and upper bounds on the cardinality of two-burst $(t_1,t_2)$-DI ECCs. 
	Based on these results,
	we propose novel constructions of binary  ECCs capable of correcting two bursts of $(t_1,t_2)$-DI for two positive integers  $t_1$ and $t_2$.
	The resulting codes  achieve substantially improved computational efficiency compared with those obtained by a direct application of the syndrome compression technique.
	As the case of $t_2=0$ has been addressed in [31], and the case of $t_1=0$ can be treated similarly,  we restrict attention to the setting $t_1,t_2>0$ throughout this paper.

	The remainder of this paper is organized as follows.
	Section \ref{sec:Preliminaries} introduces some necessary notation, definitions, and preliminary observations.
	In Section \ref{sec:bounds}, we first establish the equivalence among ECCs correcting two bursts of $(t_1,t_2)$-DI errors, two bursts of $(t_2,t_1)$-DI errors, and one burst of $(t_1,t_2)$-DI together with one burst of $(t_2,t_1)$-DI errors, and then analyze the lower and upper bounds on the cardinality of two-burst $(t_1,t_2)$-DI ECCs. 
	The constructions of two-burst $(t_1,t_2)$-DI ECCs are developed in Sections \ref{sec:syndrome} and \ref{sec:T=2}.
	Finally, Section \ref{sec:conclusions} concludes the paper.

	\section{Preliminaries}\label{sec:Preliminaries}
	
	In this section, we provide some necessary definitions and previously known results, which will be used in the subsequent constructions of two-burst $(t_1,t_2)$-DI ECCs.	For convenience,  we summarize the main notation used throughout the  paper in Table \ref{tab:notations}.
	 \begin{table}[h]
		\centering
		\caption{Main notations in this paper.}\label{tab:notations}
		\renewcommand{\arraystretch}{1.2}
			\begin{tabular}{|p{0.12\textwidth}|m{0.3\textwidth}|}
				\hline
				$[a,b]$ &$\{a,a+1,\dots,b\}$	\\
				\hline
				$[a,b)$ &$\{a,a+1,\dots,b-1\}$	\\
				\hline
				$[b]$ &$\{1,2,\dots,b\}$\\
				\hline
				$\mathbf{x}\in\Sigma_2^n$ & a binary sequence of length $n$ \\
				\hline
				 $\mathbf{x}_{\mathcal{I}}$ &   the projection of $\mathbf{x}$ onto the indices of $\mathcal{I}=\{i_1,i_2,\dots,i_j\}$ \\
				 \hline
				$X_{t-1}$, $X_{t_1-t_2}$ & matrix representations of a sequence $\mathbf{x}$ with $t-1$ rows or $t_1-t_2$ rows\\
				\hline
				$X[i]$ & the $i$-th row of $X_{t_1-t_2}$\\
				\hline
				$(t_1,t_2)$-DI &  one burst of $t_1$-deletions and one burst of $t_2$-insertions at the same component	\\
				\hline
				$(t_1,t_2)$-DS &  one burst of $t_1$-deletions and one burst of $t_2$-substitutions at the same component	\\
				\hline
				$\mathcal{E}(\mb{x})$ &
				a function capable of encoding $\mb{x}$ into a sequence with $X[1]$ being $d$-regular\\
				\hline
				$f(\mb{x})$ &a function capable of correcting two bursts of $(1,t)$-DS errors\\
				\hline
				$f_1(\mb{x})$, $f_2(\mb{x})$ &functions capable of correcting one or two bursts of $(t_1,t_2)$-DI errors, respectively\\
				\hline
				$\mathcal{B}_{m,(t_1,t_2)}^{DI}(\mathbf{x})$ & a set of sequences obtained from $\mathbf{x}$ by $m$ bursts of $(t_1,t_2)$-DI\\
				\hline
				 $\mathcal{B}_{m,(t_1,t_2)}^{DS}(\mathbf{x})$ & a set of sequences obtained from $\mathbf{x}$ by $m$ bursts of  $(t_1,t_2)$-DS\\
				\hline
				$\mathcal{B}_{(t_1,t_2),(t_2,t_1)}^{DI}(\mathbf{x})$ & the set of sequences obtained from $\mathbf{x}$ by one burst of $(t_1,t_2)$-DI together with one burst of $(t_2,t_1)$-DI\\
				\hline	
				$\mathcal{B}_{2,(t_1,t_2)}^{DI}(\mathbf{x},i_1,i_2)$ & the set of sequences that agree with $\mb{x}$ on certain substrings\\
				\hline
				$\mathcal{N}^{DI}_{2,(t_1,t_2)}(\mathbf{x})$&the set of $\mb{x}'\ne \mb{x}$ satisfying $\mathcal{B}_{2,(t_1,t_2)}^{DI}(\mathbf{x}')\cap\mathcal{B}_{2,(t_1,t_2)}^{DI}(\mathbf{x})\ne \varnothing$\\
				\hline
				$|\mathcal{N}^{DI}_{2,(t_1,t_2)}(\mathbf{x})|$ &the cardinality of $\mathcal{N}^{DI}_{2,(t_1,t_2)}(\mathbf{x})$\\
				\hline
				$\mathcal{N}^{DS}_{2,(1,t)}(\mathbf{x})$ & the set of $\mb{x}'\ne \mb{x}$ satisfying $\mathcal{B}_{2,(1,t)}^{DS}(\mathbf{x}')\cap\mathcal{B}_{2,(1,t)}^{DS}(\mathbf{x})\ne \varnothing$ \\
				\hline
				 $|\mathcal{N}^{DS}_{2,(1,t)}(\mathbf{x})|$ & the cardinality of $\mathcal{N}^{DS}_{2,(1,t)}(\mathbf{x})$\\
				\hline	
		\end{tabular}
	\end{table}
	
	For two nonnegative integers $a$ and $b$ with $a < b$, define two ordered sets $\{a,a+1,\dots,b-1\}$ and $\{a,a+1,\dots,b\}$ by $[a,b)$ and $[a,b]$, respectively.
	When $a=1$, we abbreviate $[1,b]$ as $[b]$.
	The cardinality of a set $\cal{A}$ is denoted by $|\mathcal{A}|$.
	For an $m\times n$ matrix $A$, denote the $i$-th row by $A[i],i\in[m]$.	
	For any positive integer $q\ge 2$, let $\Sigma_q=\{0,1,\dots,q-1\}$. 
	Let $|\mathbf{x}|=n$ be the length of $\mathbf{x}=(x_1,x_2,\dots,x_n)\in\Sigma_2^n$.
	For an index set $\mathcal{I}=\{i_1,i_2,\dots,i_j\}\subseteq[n]$,  $\mathbf{x}_{\mathcal{I}}=(x_{i_1},x_{i_2},\dots,x_{i_j})$ represents the projection of $\mathbf{x}$ onto the indices of $\mathcal{I}$. In this case, $\mathbf{x}_{\mathcal{I}}$ is referred to  as a \textit{substring} of $\mathbf{x}$ when $i_{j'-1}=i_{j'}-1$ for all $j'\in[2,j]$.
	If $x_i=x_{i+1}$ for every $i\in[i_1,i_j-1]$, we call the substring $(x_{i_1},x_{i_2},\dots,x_{i_j})$ a \textit{run} of $\mathbf{x}$, where $x_{i_1-1}\ne x_{i_1},x_{i_j}\ne x_{i_j+1}$, with the convention that $x_1\ne x_0,x_n\ne x_{n+1}$. If $x_i=x_{i+2}$ for every $i\in[i_1,i_j-2]$, it forms an \textit{alternating substring} in $\mathbf{x}$. 
	For example, if $\mathbf{x}=0101011110$, then the substring $1111$ forms a run of length $4$, while $010101$ forms an alternating substring of length $6$ in $\mathbf{x}$.
	
	\begin{definition}\label{def:DI}
		For a binary sequence $\mathbf{x}=(x_1,x_2,\dots,x_n)\in\Sigma_2^n$ and two nonnegative integers $t_1,t_2$,
		if the substring $(x_i,x_{i+1},\dots,x_{i+t_1-1})$ is deleted and then $(b_1,b_2,\dots,b_{t_2})$ is inserted at the $i$-th coordinate of $\mathbf{x}$, where  $b_1\ne x_i,b_{t_2}\ne x_{i+t_1-1}$ for $i\in[n-t_1+1]$, we say that \textit{one burst of $(t_1,t_2)$-DI} occurs at the $i$-th component of $\mathbf{x}$.
	\end{definition}	
	\begin{definition}\label{def:DS}
		For a binary sequence $\mathbf{x}=(x_1,x_2,\dots,x_n)\in\Sigma_2^n$ and two nonnegative integers $t_1,t_2$,
		if the substring $(x_i,x_{i+1},\dots,x_{i+t_1-1})$ is deleted and then $(x_{i+t_1},$ $x_{i+t_1+1},\dots,$ $x_{i+t_1+t_2-1})$ is substituted by $(b_1,b_2,\dots,b_{t_2})$, where  $b_1\ne x_i$ for $i\in[n-t_1-t_2+1]$, we say that \textit{one burst of $t_1$-deletions-$t_2$-substitutions ($(t_1, t_2)$-DS)} occurs at the $i$-th component of $\mathbf{x}$.  Note that when $t_2=0$, one burst of $(t_1,t_2)$-DS reduces to the case of one burst of $t_1$-deletions, i.e., there do not exist the substrings $(x_{i+t_1},$ $x_{i+t_1+1},\dots,$ $x_{i+t_1+t_2-1})$ and $(b_1,b_2,\dots,b_{t_2})$.
	\end{definition}
	
	Let $\mathcal{B}_{m,(t_1,t_2)}^{DI}(\mathbf{x})$ and $\mathcal{B}_{m,(t_1,t_2)}^{DS}(\mathbf{x})$ be the sets of sequences obtained from $\mathbf{x}$ by $m$ bursts of $(t_1,t_2)$-DI and $m$ bursts of $(t_1,t_2)$-DS, respectively. We simply write $\mathcal{B}_{(t_1,t_2)}^{DI}(\mathbf{x})$ and  $\mathcal{B}_{(t_1,t_2)}^{DS}(\mathbf{x})$ when $m=1$.

	Thus, if $\mathbf{x}'\in \mathcal{B}_{(t_1,t_2)}^{DI}(\mathbf{x})$, for $i\in[n-t_1+1]$, we can write
	\begin{align*}
		\mathbf{x}'
		=(x_1,x_2,\dots,x_{i-1},b_1,b_2,\dots,b_{t_2},x_{i+t_1},\dots,x_n).
	\end{align*}
	If $\mathbf{x}''\in \mathcal{B}_{(t_1,t_2)}^{DS}(\mathbf{x})$, for $i\in[n-t_1-t_2+1]$, we can write
	\begin{align*}
		\mathbf{x}''
		=(x_1,x_2,\dots,x_{i-1},b_1,b_2,\dots,b_{t_2},x_{i+t_1+t_2},\dots,x_n).
	\end{align*}
	
	We remark that for one burst of $(t_1, t_2)$-DS errors, it is not necessary to set $x_{i+t_1+r-1}\ne b_r$ for any $r\in[t_2]$. This means that
	the $t_2$-substitution errors do not have to be contiguous, i.e., there are exactly $t_1$-deletions and up to $t_2$-substitutions in $\mathbf{x}''$.

	Next,  we formally define ECCs capable of correcting multiple bursts of $(t_1,t_2)$-DI. It is crucial to emphasize that the model developed throughout this work assumes the presence of $m$ non-overlapping burst errors in the sequence.
	\begin{definition}
	A subset of $\Sigma_2^n$, denoted by $\mathcal{C}$, is called an \textit{$m$-burst $(t_1,t_2)$-DI ECC} if it is capable of correcting $m$ bursts of $(t_1,t_2)$-DI errors. That is, for any two distinct sequences $\mathbf{x},\mathbf{y}\in\mathcal{C}$, we have
		\begin{align*}
			\mathcal{B}_{m,(t_1,t_2)}^{DI}(\mathbf{x})\cap \mathcal{B}_{m,(t_1,t_2)}^{DI}(\mathbf{y})=\varnothing.
		\end{align*}
	\end{definition}

	
	In what follows, we define \textit{$d$-regular sequences}, which serve as a critical component  throughout the paper.
	\begin{definition}\label{def:regular}
		A sequence $\mathbf{x}\in\Sigma_2^n$ is referred to as a \textit{$d$-regular} sequence if each substring of $\mathbf{x}$ of length at least $d\log n$ contains both $00$ and $11$.
	\end{definition}

\begin{lemma}[Lemma 11, \cite{guruswami2021explicit}]\label{lem:d-regular}
	For $d\ge7$ and $\lfloor d\log n/2\rfloor n^{0.15d-1}\ge12$, the number of $d$-regular sequences of length $n$ is at least $2^{n-1}$. In particular, when $d=7$, it suffices that $n\ge9$.
\end{lemma}


In Definition \ref{def:regular}, the parameter $d$ is a constant that can be chosen appropriately. Throughout this paper, we fix $d=7$. The following lemma presents an encoding method for constructing $d$-regular sequences with only one bit of redundancy.

\begin{lemma}[Lemma 11, \cite{guruswami2021explicit}]\label{regular}
	For an integer  $M\ge 2^{n-1}$, there exists a function $\mathrm{RegEnc}:[M]\rightarrow\Sigma_2^n$, which can be computed in near-linear time with a polynomial-size lookup table, such that $\mathrm{RegEnc}(\cdot)$ is a $d$-regular sequence.
\end{lemma}

	\begin{observation}\label{observation}
		If $\mathbf{x}\in\Sigma_2^n$ is a $d$-regular sequence, then  the length of each run and
		alternating substring in $\mathbf{x}$ is at most $d\log n$ since 
		such strings cannot contain both $00$ and $11$.
	\end{observation}

	\section{Equivalence of Different Burst ECCs and Bounds on their Cardinality}\label{sec:bounds}
	
	In this section, we first establish that ECCs capable of correcting two bursts of $(t_1,t_2)$-DI are equivalent to those correcting two bursts of $(t_2,t_1)$-DI, as well as to ECCs correcting one burst of $(t_1,t_2)$-DI together with one burst of $(t_2,t_1)$-DI. Subsequently, we derive lower and upper bounds on the cardinality of two-burst $(t_1,t_2)$-DI ECCs.
	
	\subsection{The Equivalence of Different Burst ECCs}
	We characterize the equivalence of different burst ECCs in the following two theorems.
	\begin{theorem}\label{lem:t1t2-t2t1}
	 A code $\mathcal{C}$ can correct two bursts of $(t_1,t_2)$-DI if and only if it can correct two bursts of $(t_2,t_1)$-DI.
	\end{theorem}
	\begin{proof}
		We show the `only if' direction by contradiction, and note that the `if' part can be proved similarly.
		Since $\mathcal{C}$ can correct two bursts of $(t_1,t_2)$-DI, for any $\mathbf{x},\mathbf{y}\in\mathcal{C}$, we have
		$$
			\mathcal{B}_{2,(t_1,t_2)}^{DI}(\mathbf{x})\cap \mathcal{B}_{2,(t_1,t_2)}^{DI}(\mathbf{y})=\varnothing.
		$$
		
		Assume that $\mathcal{C}$ cannot correct two bursts of $(t_2,t_1)$-DI. Then, there exist two different sequences $\mathbf{x},\mathbf{y}\in\mathcal{C}$ such that
		$$
			\mathcal{B}_{2,(t_2,t_1)}^{DI}(\mathbf{x})\cap \mathcal{B}_{2,(t_2,t_1)}^{DI}(\mathbf{y})\ne\varnothing.
		$$
		That is, we can find a sequence $\mathbf{s}\in\Sigma_2^{n-2t_2+2t_1}$ such that 
		$$\mathbf{s}\in \mathcal{B}_{2,(t_2,t_1)}^{DI}(\mathbf{x})\cap \mathcal{B}_{2,(t_2,t_1)}^{DI}(\mathbf{y}).$$ 
		Suppose that $\mathbf{s}$ is obtained from $\mathbf{x}$ by deleting $(x_{i_1},x_{i_1+1},$ $\dots,x_{i_1+t_2-1})$ and $(x_{i_2},x_{i_2+1},\dots,x_{i_2+t_2-1})$, where $i_2\ge i_1+t_2$ ensures the separation between deletion positions, and subsequently inserting $(b_1,b_2,\dots,b_{t_1})$ and $(b'_1,b'_2,\dots,b'_{t_1})$ at the $i_1$-th and $i_2$-th components of the original sequence $\mathbf{x}$. Then we get
		\begin{align}\label{eqn:s-x}
			\mathbf{s}=&x_1,\dots, x_{i_1-1},b_1,b_2,\dots,b_{t_1},x_{i_1+t_2},\dots, x_{i_2-1},\nonumber\\
		&b'_1,b'_2,\dots,b'_{t_1},x_{i_2+t_2},\dots,x_n.
		\end{align}
		Meanwhile, if $\mathbf{s}$ is obtained from $\mathbf{y}$ by deleting $(y_{j_1},y_{j_1+1},$ $\dots,y_{j_1+t_2-1})$ and  $(y_{j_2},y_{j_2+1},\dots,y_{j_2+t_2-1})$, where $j_2\ge j_1+t_2$, and subsequently inserting $(c_1,c_2,\dots,c_{t_1})$ and $(c'_1,c'_2,\dots,c'_{t_1})$ at the $j_1$-th and $j_2$-th components of the original sequence $\mathbf{y}$, then we get
		\begin{align}\label{eqn:s-y}
			\mathbf{s}=&y_1,\dots, y_{j_1-1},c_1,c_2,\dots,c_{t_1},y_{j_1+t_2},\dots y_{j_2-1},\nonumber\\
			&c'_1,c'_2,\dots,c'_{t_1},y_{j_2+t_2},\dots,y_n.
		\end{align}
		
	In the following, we consider two cases according to whether the burst positions in $\mb{x}$ and $\mb{y}$ overlap. To clarify, we illustrate the two cases in Fig. \ref{fig:case}.
	\begin{figure}[t]
		\centering
		 \hspace*{-5cm}
			\begin{subfigure}{\textwidth}
				\centering
				\begin{tikzpicture}[scale=0.5, transform shape]
					\draw[-] (0,0) -- (1,0);
					\draw[-] (2.7,0) -- (4.5,0);
					\draw[-] (6.1,0) -- (15,0);

					\node at (-0.2,0) {$\mathbf{x}$};

					\node at (1,0.3) {$i_1-1$};
					\node at (1.9,0){$b_1,\dots,b_{t_1}$};
					\node at (3.1,0.3) {$i_1+t_2$};
					\node at (4.5,0.3) {$i_2-1$};
					\node at (5.4,0){$b'_1,\dots,b'_{t_1}$};
					\node at (6.5,0.3) {$i_2+t_2$};

					\fill (1,0) circle (1pt);
					\fill (2.7,0) circle (1pt);
					\fill (4.5,0) circle (1pt);
					\fill (6.1,0) circle (1pt);
					
					\draw[-] (0,-1) -- (7,-1);
					\draw[-] (8.7,-1) -- (10.5,-1);
					\draw[-] (12.2,-1) -- (15,-1);

					\node at (-0.2,-1) {$\mathbf{y}$};

					\node at (5.4,-0.7) {$i_2+2t_1-t_2$};
					\node at (7,-0.7) {$j_1-1$};
					\node at (7.9,-1){$c_1,\dots,c_{t_1}$};
					\node at (9.1,-0.7) {$j_1+t_2$};
					\node at (10.5,-0.7) {$j_2-1$};
					\node at (11.4,-1){$c'_1,\dots,c'_{t_1}$};
					\node at (12.5,-0.7) {$j_2+t_2$};
					
					\fill (6.1,-1) circle (1pt);
					\fill (7,-1) circle (1pt);
					\fill (8.7,-1) circle (1pt);
					\fill (10.5,-1) circle (1pt);
					\fill (12.2,-1) circle (1pt);

				\end{tikzpicture}
				\caption*{i) $i_1+t_2\le i_2\le j_1-2t_1+t_2\le j_2-2t_1$}
			\end{subfigure}
			
			\par
			\hspace*{-5cm}\begin{subfigure}{\textwidth}
				\centering
				
				\begin{tikzpicture}[scale=0.5, transform shape]
					\draw[-] (0,0) -- (1,0);
					\draw[-] (2.7,0) -- (4.5,0);
					\draw[-] (6.1,0) -- (15,0);

					\node at (-0.2,0) {$\mathbf{x}$};

					\node at (1,0.3) {$i_1-1$};
					\node at (1.9,0){$b_1,\dots,b_{t_1}$};
					\node at (3.1,0.3) {$i_1+t_2$};
					\node at (4.0,-0.3) {$j_1+t_2-2t_1$};\fill (4.0,0) circle (1pt);
					\draw[->] (4.0,-0.2) -- (4.0,0);
					\node at (4.5,0.3) {$i_2-1$};
					\node at (5.4,0){$b'_1,\dots,b'_{t_1}$};
					\node at (6.5,0.3) {$i_2+t_2$};
					\node at (7.3,-0.3) {$j_1+2t_2-t_1$};
					\draw[->] (6.7,-0.2) -- (6.7,0);
					
					\fill (1,0) circle (1pt);
					\fill (2.7,0) circle (1pt);
					\fill (4.5,0) circle (1pt);
					\fill (6.1,0) circle (1pt);
					\fill (6.7,0) circle (1pt);
					
					\draw[-] (0,-1) -- (5,-1);
					\draw[-] (6.7,-1) -- (8.5,-1);
					\draw[-] (10.2,-1) -- (15,-1);

					\node at (-0.2,-1) {$\mathbf{y}$};
					
					\node at (4.0,-1.3) {$j_1-t_1$};\fill (4.0,-1) circle (1pt);
					\draw[->] (3.7,-1.2) -- (4.0,-1);
					\node at (3.5,-0.7) {$i_2+t_1-t_2-1$};\fill (4.5,-1) circle (1pt);
					\draw[->] (4.1,-0.8) -- (4.5,-1);
					\node at (5.2,-0.7) {$j_1-1$};
					\node at (5.9,-1){$c_1,\dots,c_{t_1}$};
					\node at (7.1,-0.7) {$j_1+t_2$};
					\node at (8.5,-0.7) {$j_2-1$};
					\node at (9.4,-1){$c'_1,\dots,c'_{t_1}$};
					\node at (10.5,-0.7) {$j_2+t_2$};

					\fill (5,-1) circle (1pt);
					\fill (6.7,-1) circle (1pt);
					\fill (8.5,-1) circle (1pt);
					\fill (10.2,-1) circle (1pt);

				\end{tikzpicture}
				\caption*{ii) $i_1+t_2\le i_2\le j_1-t_1+t_2\le j_2-t_1~(j_1 < i_2+2t_1-t_2)$}
			\end{subfigure}
			\caption{The cases i) and ii).}
			\label{fig:case}
		\end{figure}

\textbf{Case i) $i_1+t_2\le i_2\le j_1-2t_1+t_2\le j_2-2t_1$}. By comparing \eqref{eqn:s-x} and \eqref{eqn:s-y}, we obtain
	\begin{align}\label{eqn:xy}
		&	(x_1,\dots,x_{i_1-1})=(y_1,\dots,y_{i_1-1}),\notag\\
		&	(b_1,\dots,b_{t_1})=(y_{i_1},\dots,y_{i_1+t_1-1}),\notag\\
		&	(x_{i_1+t_2},\dots,x_{i_2-1})=(y_{i_1+t_1},\dots,y_{i_2+t_1-t_2-1}),\notag\\
		&	(b'_1,\dots,b'_{t_1})=(y_{i_2+t_1-t_2},\dots,y_{i_2+2t_1-t_2-1}),\\
		&	(x_{i_2+t_2},\dots,x_{j_1+2t_2-2t_1-1})=(y_{i_2+2t_1-t_2},\dots,y_{j_1-1}),\notag\\
		&	(x_{j_1+2t_2-2t_1},\dots,x_{j_1+2t_2-t_1-1})=(c_1,\dots,c_{t_1}),\notag\\
		&	(x_{j_1+2t_2-t_1},\dots,x_{j_2+t_2-t_1-1})=(y_{j_1+t_2},\dots,y_{j_2-1}),\notag\\
		&	(x_{j_2+t_2-t_1},\dots,x_{j_2+t_2-1})=(c'_1,\dots,c'_{t_1}),\notag\\
		&	(x_{j_2+t_2},\dots,x_n)=(y_{j_2+t_2},\dots,y_n).\notag
	\end{align}
	Therefore, by deleting $
	(x_{j_1+2t_2-2t_1},\dots,x_{j_1+2t_2-t_1-1})$ and 
	$
	(x_{j_2+t_2-t_1},\dots,x_{j_2+t_2-1})
	$
	from $\mathbf{x}$, then inserting
	$
	(y_{j_1},\dots,$ $y_{j_1+t_2-1})
	$
	and 
	$
	(y_{j_2},\dots,y_{j_2+t_2-1})
	$ at the $j_1+2t_2-2t_1$-th and $j_2+t_2-t_1$-th components of $\mathbf{x}$, respectively, we have
	\begin{align*}
		\mathbf{x}'=	(x_1,\dots,x_{j_1+2t_2-2t_1-1},y_{j_1},\dots,y_{j_1+t_2-1},x_{j_1+2t_2-t_1},\notag\\
		\dots,x_{j_2+t_2-t_1-1},y_{j_2},\dots,y_{j_2+t_2-1},x_{j_2+t_2},\dots,x_n).
	\end{align*}
	By deleting $
	(y_{i_1},\dots,y_{i_1+t_1-1})
	$ and 
	$
	(y_{i_2+t_1-t_2},\dots,$ $y_{i_2+2t_1-t_2-1})
	$
	from $\mathbf{y}$, then inserting $(x_{i_1},\dots,x_{i_1+t_2-1})$ and $(x_{i_2},\dots,x_{i_2+t_2-1})$ at the $i_1$-th and $i_2+t_1-t_2$-th components of $\mathbf{y}$, respectively, we obtain
	\begin{align*}
		\mathbf{y}'=&	(y_1,\dots,y_{i_1-1},x_{i_1},\dots,x_{i_1+t_2-1},y_{i_1+t_1},\dots,\\
		&y_{i_2+t_1-t_2-1},	x_{i_2},\dots,x_{i_2+t_2-1},y_{i_2+2t_1-t_2},\dots,y_n).
	\end{align*}
	
	With the equalities  in \eqref{eqn:xy}, $\mathbf{x}'=\mathbf{y}'$ follows.
	This implies that 
	$$\mathcal{B}_{2,(t_1,t_2)}^{DI}(\mathbf{x})\cap \mathcal{B}_{2,(t_1,t_2)}^{DI}(\mathbf{y})\ne \varnothing,$$
	 which contradicts the fact that $\mathcal{C}$ is an ECC for correcting two bursts of $(t_1,t_2)$-DI.
	
	\textbf{Case ii) $i_1+t_2\le i_2\le j_1-t_1+t_2\le j_2-t_1~(j_1 < i_2+2t_1-2t_2)$}. As shown in Fig. 1-ii), in this case, the second burst error in $\mb{x}$ intersects with the first one in $\mb{y}$. By comparing  (1) and (2), we obtain 
	\begin{align}\label{eqn:xy1}
		\begin{array}{lll}
			(x_1,\dots,x_{i_1-1})=(y_1,\dots,y_{i_1-1}),\\
			(b_1,\dots,b_{t_1})=(y_{i_1},\dots,y_{i_1+t_1-1}),\\
			(x_{i_1+t_2},\dots,x_{i_2-1})=(y_{i_1+t_1},\dots,y_{i_2+t_1-t_2-1}),\\
			(x_{j_1+2t_2-t_1},\dots,x_{j_2+t_2-t_1-1})=(y_{j_1+t_2},\dots,y_{j_2-1}),\\
			(x_{j_2+t_2-t_1},\dots,x_{j_2+t_2-1})=(c'_1,\dots,c'_{t_1}),\\
			(x_{j_2+t_2},\dots,x_n)=(y_{j_2+t_2},\dots,y_n).\\
		\end{array}
	\end{align}

	To obtain a common sequence such that 
		$$\mathcal{B}_{2,(t_1,t_2)}^{DI}(\mathbf{x})\cap\mathcal{B}_{2,(t_1,t_2)}^{DI}(\mathbf{y})\ne\varnothing,$$ it is essential to consider the overlapping part of two bursts in Fig. 1-ii), i.e., $(b_1',\dots,b_{t_1}',x_{i_2+t_2},\dots,x_{j_1+2t_2-t_1-1})$ and $(y_{i_2+t_1-t_2},\dots,y_{j_1-1}, c_1,\dots,c_{t_1})$.
		Since the total number of symbols to be deleted from two bursts is $2t_1$, we select the substrings $(x_{j_1+t_2-2t_1},\dots,x_{j_1+t_2-t_1-1})$ in $\mb{x}$ and $(y_{j_1+t_2-t_1},\dots,y_{j_1+t_2-1})$
		in $\mathbf{y}$. 
		These two substrings together completely cover the intersection region and have a combined length of $2t_1$. We emphasize that the deleted substrings are not unique, and the above selection represents only a feasible realization.

	Therefore, by deleting $
	(x_{j_1+t_2-2t_1},\dots,x_{j_1+t_2-t_1-1})
	$ and $
	(x_{j_2+t_2-t_1},\dots,x_{j_2+t_2-1})
	$  from $\mathbf{x}$, then inserting $(y_{j_1-t_1},\dots,y_{j_1+t_2-t_1-1})$  and $(y_{j_2},\dots,y_{j_2+t_2-1})$ at the $j_1+t_2-2t_1$-th and $j_2+t_2-t_1$-th components of $\mathbf{x}$, respectively, we can obtain
	\begin{align*}
		\mathbf{x}''=	(&x_1,\dots,x_{j_1+t_2-2t_1-1},y_{j_1-t_1},\dots,y_{j_1+t_2-t_1-1},\\&x_{j_1+t_2-t_1},\dots,
		x_{j_2+t_2-t_1-1},y_{j_2},\dots,y_{j_2+t_2-1},\\
		&x_{j_2+t_2},\dots,x_n).
	\end{align*}
	By deleting $
	(y_{i_1},\dots,y_{i_1+t_1-1})$ and 
	$
	(y_{j_1+t_2-t_1},\dots,y_{j_1+t_2-1})
	$ 
	from $\mathbf{y}$, then inserting $(x_{i_1},\dots,x_{i_1+t_2-1})$ and $(x_{j_1+t_2-t_1},$ $\dots,x_{j_1+2t_2-t_1-1})$ at the $i_1$-th and $j_1+t_2-t_1$-th components of $\mathbf{y}$, respectively, we can obtain
	\begin{align*}
		\mathbf{y}''=	(&y_1,\dots,y_{i_1-1},x_{i_1},\dots,x_{i_1+t_2-1},y_{i_1+t_1},\dots,\\
		&y_{j_1+t_2-t_1-1},
		x_{j_1+t_2-t_1},\dots,x_{j_1+2t_2-t_1-1},y_{j_1+t_2},\\&\dots,y_n).
	\end{align*}
	
	Since $i_2+t_1-t_2\le j_1< i_2+2t_1-2t_2$, we know $(x_{i_1+t_2},\dots,$ $x_{j_1+t_2-2t_1})=(y_{i_1+t_1},\dots,y_{j_1-t_1})$.
	With the equalities in \eqref{eqn:xy1}, $\mathbf{x}''=\mathbf{y}''$ holds.
	This implies that $$\mathcal{B}_{2,(t_1,t_2)}^{DI}(\mathbf{x})\cap \mathcal{B}_{2,(t_1,t_2)}^{DI}(\mathbf{y})\ne \varnothing,$$ which contradicts the fact that $\mathcal{C}$ is an ECC for correcting two bursts of $(t_1,t_2)$-DI.
	
We note that additional cases may arise depending on the relative positions of two bursts in $\mathbf{x}$ and $\mathbf{y}$, for example, $i_1+t_1\le j_1\le i_2-t_2\le j_2-t_1-t_2$, or $i_1+t_1\le j_1\le j_2-t_2\le i_2-t_1-t_2$, and so on. These cases can be proved analogously to Cases i) and ii).
		By comparing \eqref{eqn:s-x} and \eqref{eqn:s-y} under these configurations, we can establish the corresponding equivalence among the symbols,  similar to those in \eqref{eqn:xy} and \eqref{eqn:xy1}. Based upon this, by first deleting and then inserting two substrings into $\mathbf{x}$ and $\mathbf{y}$, we can obtain $$\mathcal{B}_{2,(t_1,t_2)}^{DI}(\mathbf{x})\cap \mathcal{B}_{2,(t_1,t_2)}^{DI}(\mathbf{y})\ne \varnothing,$$ 
		which leads to a contradiction.
				This completes the proof.
	\end{proof}

	\begin{theorem}\label{lem:t1t2+t2t1}
		A code $\mathcal{C}$ can correct two bursts of $(t_1,t_2)$-DI if and only if it can correct one burst of $(t_1,t_2)$-DI together with one burst of $(t_2,t_1)$-DI.
	\end{theorem}
	\begin{proof}
		The proof follows a similar strategy to that of Theorem \ref{lem:t1t2-t2t1}.
		Since $\mathcal{C}$ can correct two bursts of $(t_1,t_2)$-DI, for any $\mathbf{x},\mathbf{y}\in\mathcal{C}$, we have
		\begin{align*}
			\mathcal{B}_{2,(t_1,t_2)}^{DI}(\mathbf{x})\cap \mathcal{B}_{2,(t_1,t_2)}^{DI}(\mathbf{y})=\varnothing.
		\end{align*}
		
		Assume that $\mathcal{C}$ cannot correct one burst of $(t_1,t_2)$-DI together with one burst of $(t_2,t_1)$-DI. Then, there exist two different sequences $\mathbf{x},\mathbf{y}\in\mathcal{C}$ such that
		\begin{align*}
			\mathcal{B}_{(t_1,t_2),(t_2,t_1)}^{DI}(\mathbf{x})\cap \mathcal{B}_{(t_1,t_2),(t_2,t_1)}^{DI}(\mathbf{y})\ne\varnothing,
		\end{align*}
		where $\mathcal{B}_{(t_1,t_2),(t_2,t_1)}^{DI}(\mathbf{x})$ represents the set of sequences obtained from $\mathbf{x}$ by  one burst of $(t_1,t_2)$-DI together with one burst of $(t_2,t_1)$-DI.
		That is, we can find a sequence $\mathbf{s}\in\Sigma_2^{n}$ such that  
		$$\mathbf{s}\in \mathcal{B}_{(t_1,t_2),(t_2,t_1)}^{DI}(\mathbf{x})\cap \mathcal{B}_{(t_1,t_2),(t_2,t_1)}^{DI}(\mathbf{y}).$$ 
		Suppose that $\mathbf{s}$ is obtained from $\mathbf{x}$ by deleting $(x_{i_1},x_{i_1+1},\dots,x_{i_1+t_1-1})$ and $(x_{i_2},x_{i_2+1},\dots,x_{i_2+t_2-1})$, where $i_2\ge i_1+t_1$, and inserting $(b_1,b_2,\dots,b_{t_2})$ and $(b'_1,b'_2,\dots,b'_{t_1})$ at the $i_1$-th and $i_2$-th components of $\mathbf{x}$. Then we have
		\begin{align}\label{eqn:s-x2}
			\mathbf{s}=&x_1,\dots, x_{i_1-1},b_1,b_2,\dots,b_{t_2},x_{i_1+t_1},\dots, x_{i_2-1},\notag\\
			&b'_1,b'_2,\dots,b'_{t_1},x_{i_2+t_2},\dots,x_n.
		\end{align}
		Meanwhile, if $\mathbf{s}$ is obtained from $\mathbf{y}$ by deleting $(y_{j_1},y_{j_1+1},$ $\dots,y_{j_1+t_1-1})$ and $(y_{j_2},y_{j_2+1},\dots,y_{j_2+t_2-1})$, where $j_2\ge j_1+t_1$,
		and inserting $(c_1,c_2,\dots,c_{t_2})$ and $(c'_1,c'_2,\dots,c'_{t_1})$ at the $j_1$-th and $j_2$-th components of $\mathbf{y}$, then we have
		\begin{align}\label{eqn:s-y2}
			\mathbf{s}=&y_1,\dots, y_{j_1-1},c_1,c_2,\dots,c_{t_2},y_{j_1+t_1},\dots, y_{j_2-1},\notag\\
			&c'_1,c'_2,\dots,c'_{t_1},y_{j_2+t_2},\dots,y_n.
		\end{align}	
		
		To illustrate the main argument, we consider the ordering $i_1+t_1\le i_2\le j_1-t_2\le j_2-t_1-t_2$. By comparing  \eqref{eqn:s-x2} and \eqref{eqn:s-y2}, we obtain
		\begin{align}\label{eqn:xy2}
			\begin{array}{lll}
				(x_1,\dots,x_{i_1-1})=(y_1,\dots,y_{i_1-1}),\\
				(b_1,\dots,b_{t_2})=(y_{i_1},\dots,y_{i_1+t_2-1}),\\
				(x_{i_1+t_1},\dots,x_{i_2-1})=(y_{i_1+t_2},\dots,y_{i_2+t_2-t_1-1}),\\
				(b'_1,\dots,b'_{t_1})=(y_{i_2+t_2-t_1},\dots,y_{i_2+t_2-1}),\\
				(x_{i_2+t_2},\dots,x_{j_1-1})=(y_{i_2+t_2},\dots,y_{j_1-1}),\\
				(x_{j_1},\dots,x_{j_1+t_2-1})=(c_1,\dots,c_{t_2}),\\
				(x_{j_1+t_2},\dots,x_{j_2+t_2-t_1-1})=(y_{j_1+t_1},\dots,y_{j_2-1}),\\
				(x_{j_2-t_1+t_2},\dots,x_{j_2+t_2-1})=(c'_1,\dots,c'_{t_1}),\\
				(x_{j_2+t_2},\dots,x_n)=(y_{j_2+t_2},\dots,y_n).\\
			\end{array}
		\end{align}
		Therefore, by deleting 
		$
			(x_{i_1},x_{i_1+1},\dots,x_{i_1+t_1-1})
		$ and 
		$
			(x_{j_2+t_2-t_1},\dots,x_{j_2+t_2-1})
			$  from $\mathbf{x}$, and then inserting $(y_{i_1},\dots,y_{i_1+t_2-1})$  and $(y_{j_2},\dots,y_{j_2+t_2-1})$ at the $i_1$-th and $j_2+t_2-t_1$-th components of $\mathbf{x}$, respectively, we can obtain
		\begin{align*}
			\mathbf{x}'''=	(&x_1,\dots,x_{i_1-1},y_{i_1},\dots,y_{i_1+t_2-1},x_{i_1+t_1},\dots,\\
			&x_{j_2+t_2-t_1-1},y_{j_2},\dots,y_{j_2+t_2-1},x_{j_2+t_2},\dots,x_n).
		\end{align*}
		By deleting $
			(y_{i_2+t_2-t_1},\dots,y_{i_2+t_2-1})
		$ and $
			(y_{j_1},\dots,y_{j_1+t_1-1})$ from $\mathbf{y}$, and then inserting $(x_{i_2},\dots,x_{i_2+t_2-1})$ and $(x_{j_1},\dots,$ $x_{j_1+t_2-1})$ at the $i_2+t_2-t_1$-th and $j_1$-th components of $\mathbf{y}$, respectively, we obtain
		\begin{align*}
			\mathbf{y}'''=	(&y_1,\dots,y_{i_2+t_2-t_1-1},x_{i_2},\dots,x_{i_2+t_2-1},y_{i_2+t_2},\dots,\\
			&y_{j_1-1},x_{j_1},\dots,x_{j_1+t_2-1},y_{j_1+t_1},\dots,y_n).
		\end{align*}

		From the equalities in \eqref{eqn:xy2}, it follows that $\mathbf{x}'''=\mathbf{y}'''$.
		This implies that 
		$$\mathcal{B}_{2,(t_1,t_2)}^{DI}(\mathbf{x})\cap \mathcal{B}_{2,(t_1,t_2)}^{DI}(\mathbf{y})\ne \varnothing,$$ which contradicts the fact that $\mathcal{C}$ is an ECC for correcting two bursts of $(t_1,t_2)$-DI.
		
		 The proof of the case where $\mathbf{s}$ is obtained from $\mathbf{y}$ by deleting 
		 $
		 	(y_{j_1},y_{j_1+1},\dots,y_{j_1+t_2-1})
		 $ and $
	 	(y_{j_2},y_{j_2+1},\dots,y_{j_2+t_1-1})
	 $ with $j_2\ge j_1+t_2$,
		and inserting $(c_1,c_2,\dots,c_{t_1})$ and $(c'_1,c'_2,\dots,c'_{t_2})$ at the $j_1$-th and $j_2$-th components of $\mathbf{y}$, respectively, is analogous to the preceding case. This completes the proof.
	\end{proof}
	
	The following corollary is immediate from an argument similar to
		those of  Theorems \ref{lem:t1t2-t2t1} and  \ref{lem:t1t2+t2t1}.
		\begin{corollary}
		Let $t_1$, $t_2$ be two nonnegative integers with $t_1\ge t_2$. A code $\mathcal{C}$ can correct $m$ bursts of $(t_1,t_2)$-DI if and only if it can correct $m_1$ bursts of $(t_1,t_2)$-DI together with $m_2$ bursts of $(t_2,t_1)$-DI provided that $m_1+m_2=m$.
	\end{corollary}

	\subsection{The Bounds on the Cardinality of Two-Burst $(t_1,t_2)$-DI ECCs}
	In this subsection, we develop both lower and upper bounds on the size of ECCs that can correct two bursts of $(t_1,t_2)$-DI errors. The lower bound is an extension of ECCs correcting two bursts of exactly $t_1$-deletion proposed in \cite{ye2024codes}, corresponding to two-burst  $(t_1,0)$-DI ECCs. On the other hand, the  upper bound extends the result in \cite{sun2024asymptotically}, which considers one-burst $(t_1,t_2)$-DI ECCs.

	We denote the cardinality of a two-burst $(t_1,t_2)$-DI ECC $\mathcal{C}$ of length $n$ satisfying $n> \max(2t_1,2t_2)$ by $|\mathcal{C}|$. 
		\begin{theorem}[Lower bound on the code size]\label{thm:lowB}
		For any $t_1$ and $t_2$, there exists an ECC $\mathcal{C}\subseteq\Sigma_2^n$ correcting two bursts of $(t_1,t_2)$-DI errors with
		\begin{align*}
				|\mathcal{C}|\ge\frac{2^{n}}{\min(\gamma_1,\gamma_2)},
			\end{align*}
			where 
			\begin{align*}
				\gamma_1=2^{2(t_1 + t_2)}\binom{n - 2t_1 + 2}{2} \binom{n- 2t_1+ 2t_2+1}{2}
			\end{align*}
			and 
			\begin{align*}
				\gamma_2=2^{2(t_1 + t_2)}\binom{n - 2t_2 + 2}{2} \binom{n- 2t_2+ 2t_1+1}{2}.
		\end{align*}
	\end{theorem}	
	\begin{proof}
		We define a graph $\mathcal{G} = (\mathcal{V},\mathcal{E})$ with vertex set $\mathcal{V} = \Sigma_2^n$, where two distinct vertices $\mathbf{v}, \mathbf{v}' \in \mathcal{V}$ are connected by an edge $(\mathbf{v}, \mathbf{v}') \in \mathcal{E}$ if and only if 
		\[
		\mathcal{B}_{2,(t_1,t_2)}^{DI}(\mathbf{v}) \cap \mathcal{B}_{2,(t_1,t_2)}^{DI}(\mathbf{v}') \ne \varnothing.
		\]
		This adjacency relation is denoted compactly as $\mathbf{v} \sim \mathbf{v}'$.  
		
		The independence number of $\mathcal{G}$, denoted by $N(\mathcal{V},\mathcal{E})$, represents the maximum cardinality of an independent set in $\mathcal{G}$. A code $\mathcal{C} \subseteq \Sigma_2^n$ is an ECC capable of correcting two bursts of $(t_1,t_2)$-DI if and only if $\mathcal{C}$ forms an independent set in $\mathcal{G}$. Consequently, the maximum code size satisfies $|\mathcal{C}| = N(\mathcal{V},\mathcal{E})$.  
		
		For a given vertex $\mathbf{x} \in \mathcal{V}$, let $\deg(\mathbf{x})$ denote its degree in $\mathcal{G}$, i.e., the number of vertices $\mathbf{y} \in \mathcal{V}$ such that $\mathbf{x} \sim \mathbf{y}$. Applying the probabilistic method from Theorem 1 on page 100 of \cite{alon2016probabilistic}, we derive the following lower bound on the code size,
		\begin{align*}
			|\mathcal{C}| \geq \sum_{\mathbf{x} \in \Sigma_2^n} \frac{1}{\deg(\mathbf{x}) + 1}.
		\end{align*}

		To enumerate the compatible pairs $(\mathbf{x},\mathbf{y})$, we observe that any $\mathbf{y}$ satisfying $\mathcal{B}_{2,(t_1,t_2)}^{DI}(\mathbf{x}) \cap \mathcal{B}_{2,(t_1,t_2)}^{DI}(\mathbf{y}) \ne \varnothing$ can be obtained via the following combinatorial operations:  
		(1) Delete two non-overlapping substrings of length $t_1$ from $\mathbf{x}$, yielding an intermediate string $\mathbf{x}' \in \Sigma_2^{n-2t_1}$, and subsequently insert two substrings of length $t_2$ at the deleted positions, resulting in $\mathbf{x}'' \in \Sigma_2^{n-2t_1 + 2t_2}$.  
		(2) Insert two substrings of length $t_1$ into $\mathbf{x}''$ and then delete two substrings of length $t_2$ immediately preceding the inserted positions to obtain $\mathbf{y}$.  
		This argument yields the upper bound
		\begin{align*}
			\deg(\mathbf{x}) + 1 &\leq \binom{n - 2t_1 + 2}{2} \binom{n- 2t_1+ 2t_2+1}{2} 2^{2(t_1 + t_2)}\\
			&=\gamma_1.
		\end{align*}

		Moreover, based on Theorem \ref{lem:t1t2-t2t1}, $\mathcal{C}$ is also an ECC correcting two bursts of $(t_2,t_1)$-DI errors, thus we have
			\begin{align*}
				\deg(\mathbf{x}) + 1 &\leq \binom{n - 2t_2 + 2}{2} \binom{n- 2t_2+ 2t_1+1}{2} 2^{2(t_1 + t_2)}\\
				&=\gamma_2.
			\end{align*}
			
			The equations above imply that
			\begin{align*}
				|\mathcal{C}|\ge\frac{2^{n}}{\min(\gamma_1,\gamma_2)},
			\end{align*}
		thus completing the proof.  
	\end{proof}

	\begin{theorem}[Upper bound on the code size]\label{thm:upB}
		For any two nonnegative integers $t_1$ and $t_2$, the cardinality of an ECC capable of correcting two bursts of $(t_1,t_2)$-DI satisfies 
		\begin{align*}
		|\mathcal{C}|\le {2^{n}\over{\max(a_1,a_2)}},
		\end{align*}
	where 
	\begin{align*}
		a_1=2^{2t_1-2}\left({n-2t_1+2\choose 2}+1\right),
	\end{align*}
	and
		\begin{align*}
		a_2=2^{2t_2-2}\left({n-2t_2+2\choose 2}+1\right).
	\end{align*}
	
	\end{theorem}	
	\begin{proof}
		For any $\mathbf{x}\in\mathcal{C}$, assume that $\mathbf{x}'$ and $\mathbf{x}''$ are obtained from $\mathbf{x}$ by two bursts of $(t_1,t_2)$-DI at the $i_1$-th, $i_2$-th and $i'_1$-th, $i'_2$-th
		components of $\mathbf{x}$, respectively, where $1\le i_1\le i_2-t_1\le n-2t_1+1,1\le i'_1\le i'_2-t_1\le n-2t_1+1$ and $\{i_1,i_2\}\ne \{i'_1,i'_2\}$.
		From the definition of  one burst of $(t_1,t_2)$-DI, we have 
		\begin{align*}
			x'_{i_1}\ne x_{i_1},x'_{i_2+t_2-t_1}\ne x_{i_2},
		\end{align*} 
	and 
	\begin{align*}
		x''_{i'_1}\ne x_{i'_1},x''_{i'_2+t_2-t_1}\ne x_{i'_2}.
	\end{align*}
		Without loss of generality, let $i_1\le i'_1$ and $i_2\le i'_2$. 
		Then, we can obtain $\mathbf{x}'\ne \mathbf{x}''$ since $x''_{i_1}=x_{i_1}$ when $i_1<i'_1$ or since $x''_{i_2+t_2-t_1}=x_{i_2}$ when $i_1=i'_1$ and $i_2<i'_2$.
		Consequently, for different index pairs $\{i_1,i_2\}\ne \{i'_1,i'_2\}$, the resulting sequences are not identical.
		
		Define
		\begin{align*}
			&\mathcal{B}_{2,(t_1,t_2)}^{DI}(\mathbf{x},i_1,i_2)\triangleq\\
			&\left\{\begin{array}{lll}
				\big\{\mathbf{x}'\in\Sigma_2^{n-2t_1+2t_2}:\mathbf{x}'_{[n-2t_1+2t_2]\backslash (\mathcal{I}'_1\cup\mathcal{I}'_2)}=	\mathbf{x}_{[n]\backslash (\mathcal{I}_1\cup\mathcal{I}_2)}, \\
				\hspace{2.6cm}\text{~and~}x'_{i_1}\ne x_{i_1},x'_{i_2-t_1+t_2}\ne x_{i_2}\big\},\\
					\hspace{2.4cm}\text{~if~} 1\le i_1\le i_2-t_1\le n-2t_1+1,\\
				\big\{\mathbf{x}'\in\Sigma_2^{n-2t_1+2t_2}:\mathbf{x}'_{[n-2t_1+1]}=	\mathbf{x}_{[n-2t_1+1]}\big\},\\ 
				\hspace{2.4cm} \text{~if~}  i_1= n-2t_1+2,i_2=n-t_1+2,
			\end{array}\right.
		\end{align*}
		where $\mathcal{I}'_1=[i_1,i_1+t_2-1],\mathcal{I}'_2=[i_2-t_1+t_2,i_2-t_1+2t_2-1],\mathcal{I}_1=[i_1,i_1+t_1-1]$ and $\mathcal{I}_2=[i_2,i_2+t_1-1]$.

		For any $1\le i_1\le i_2-t_1\le n-2t_1+2,1\le i'_1\le i'_2-t_1\le n-2t_1+2$ with $\{i_1,i_2\}\ne \{i'_1,i'_2\}$, it is obvious that 
		\begin{align*}
			\mathcal{B}_{2,(t_1,t_2)}^{DI}(\mathbf{x},i_1,i_2)\subseteq \mathcal{B}_{2,(t_1,t_2)}^{DI}(\mathbf{x})
		\end{align*}
		and  
		\begin{align*}
			\mathcal{B}_{2,(t_1,t_2)}^{DI}(\mathbf{x},i_1,i_2)\cap \mathcal{B}_{2,(t_1,t_2)}^{DI}(\mathbf{x},i'_1,i'_2)=\varnothing.
		\end{align*}
		Then if 
		\begin{align}\label{eqn:bi12}
			\mathcal{B}_{2,(t_1,t_2)}^{DI}(\mathbf{x})\subseteq\bigcup_{i_1=1}^{n-2t_1+2}\bigcup_{i_2=i_1+t_1}^{n-t_1+2}\mathcal{B}_{2,(t_1,t_2)}^{DI}(\mathbf{x},i_1,i_2),
		\end{align}
		we can state that $\mathcal{B}_{2,(t_1,t_2)}^{DI}(\mathbf{x},i_1,i_2)$ with $i_1\in[n-2t_1+2],i_2\in[i_1+t_1,n-t_1+2]$
	form a partition of $\mathcal{B}_{2,(t_1,t_2)}^{DI}(\mathbf{x})$.

		In what follows, we prove that the inclusion in \eqref{eqn:bi12} holds.
		Indeed, for any $\mathbf{x}'\in \mathcal{B}_{2,(t_1,t_2)}^{DI}(\mathbf{x})$, if there exist two indices $i_1\in[n-2t_1+1]$ and $i_2\in[i_1+t_1,n-t_1+1]$ such that $\mathbf{x}'_{[n-2t_1+1]}\ne	\mathbf{x}_{[n-2t_1+1]}$, then $\mathbf{x}'\in \mathcal{B}_{2,(t_1,t_2)}^{DI}(\mathbf{x},i_1,i_2)$, otherwise,
		$\mathbf{x}'\in \mathcal{B}_{2,(t_1,t_2)}^{DI}(\mathbf{x},n-2t_1+2,n-t_1+2)$.

		Therefore,
				\begin{align}\label{eqn:equ}
					\scalebox{0.9}{$
						\begin{aligned}
			\left|\mathcal{B}_{2,(t_1,t_2)}^{DI}(\mathbf{x})\right|=&\sum_{i_1=1}^{n-2t_1+1}\sum_{i_2=i_1+t_1}^{n-t_1+1}\left|\mathcal{B}_{2,(t_1,t_2)}^{DI}(\mathbf{x},i_1,i_2)\right|\\
			&+\left|\mathcal{B}_{2,(t_1,t_2)}^{DI}(\mathbf{x},n-2t_1+2,n-t_1+2)\right|\\
			=&2^{2t_2-2}\sum_{i_1=1}^{n-2t_1+1}(n-2t_1-i_1+2)+2^{2t_2-2}\\
			=&2^{2t_2-2}\left({n-2t_1+2\choose 2}+1\right).
		\end{aligned}
	$}
\end{align}
		
		Since $\mathcal{C}$ is an ECC correcting two bursts of $(t_1,t_2)$-DI errors, we have
		\begin{align*}
			&\left|\bigcup_{\mathbf{x}\in\mathcal{C}}\mathcal{B}_{2,(t_1,t_2)}^{DI}(\mathbf{x})\right|=\sum_{\mathbf{x}\in\mathcal{C}}\left|\mathcal{B}_{2,(t_1,t_2)}^{DI}(\mathbf{x})\right|\\
			=&\left|\mathcal{C}\right|\cdot 2^{2t_2-2}\left({n-2t_1+2\choose 2}+1\right)\\
			\le& 2^{n-2t_1+2t_2}.
		\end{align*}
	
	Moreover, based on Theorem \ref{lem:t1t2-t2t1}, $\mathcal{C}$ is also an ECC correcting two bursts of $(t_2,t_1)$-DI errors, thus we have
	\begin{align*}
		&\left|\bigcup_{\mathbf{x}\in\mathcal{C}}\mathcal{B}_{2,(t_2,t_1)}^{DI}(\mathbf{x})\right|=\sum_{\mathbf{x}\in\mathcal{C}}\left|\mathcal{B}_{2,(t_2,t_1)}^{DI}(\mathbf{x})\right|\\
		=&\left|\mathcal{C}\right|\cdot2^{2t_1-2}\left({n-2t_2+2\choose 2}+1\right)\\
		\le &2^{n-2t_2+2t_1}.
	\end{align*}

	Denoting $a_1=2^{2t_1-2}\left({n-2t_1+2\choose 2}+1\right)$ and $a_2=2^{2t_2-2}\left({n-2t_2+2\choose 2}+1\right)$, the equations above imply that
		\begin{align*}
			|\mathcal{C}|\le\frac{2^{n}}{\max(a_1,a_2)},
		\end{align*}
		completing the proof. 
	\end{proof}
	
	Based on Theorems \ref{thm:lowB} and \ref{thm:upB}, there exists an ECC $\mathcal{C}\subseteq\Sigma_2^n$ capable of correcting two bursts of $(t_1,t_2)$-DI errors with redundancy
	\begin{align*}
		\log (\max(a_1,a_2))\le ~r(\mathcal{C})
		\le \log\big(\min(\gamma_1,\gamma_2)\big).
	\end{align*}
	
	Furthermore, we have the following corollary.
	\begin{corollary}
		There exists an ECC $\mathcal{C}\subseteq\Sigma_2^n$ capable of correcting two bursts of $(t_1,t_2)$-DI errors with redundancy $r(\mathcal{C})$ satisfying $2\log n+O(1)\le r(\mathcal{C})\le 4\log n +o(\log n)$.
	\end{corollary}

	\begin{remark}
		When $t_1=t_2=1$, two bursts of $(1,1)$-DI coincide with exactly two substitution errors. In this case, the Hamming ball of size ${n\choose 2}+1$ is obtained, which is consistent with  Equation \eqref{eqn:equ}.
\end{remark}

	\begin{remark}
	The upper and lower bounds on the code size of two-burst $(t_1,t_2)$-DI ECCs can be naturally  extended to the case of $m$ bursts with $m\ge 2$. That is, there exists a code $\mathcal{C}\subseteq\Sigma_2^n$ capable of correcting $m$ bursts of $(t_1,t_2)$-DI errors with redundancy $r(\mathcal{C})$ satisfying $m\log n+O(1)\le r(\mathcal{C})\le 2m\log n +o(\log n)$.
	\end{remark}

	\section{Two-burst $(t_1,t_2)$-DI ECCs by Syndrome Compression Techniques}\label{sec:syndrome}
	In view of the equivalence results established in Section \ref{sec:bounds}, in what follows, we  restrict our attention to two bursts of  $(t_1,t_2)$-DI errors with $t_1\ge t_2$. Under this setting, we develop constructions of two-burst $(t_1,t_2)$-DI ECCs by using the syndrome compression technique  proposed in \cite{sima2020syndrome} in this section. 
	Although the applicability of the syndrome compression technique to the construction of such codes is relatively straightforward, we provide the detailed construction here to lay the groundwork for the next section and to facilitate a comparison in terms of computational efficiency.
	
	Before presenting the construction, we first review a function capable of correcting multiple deletions, insertions and substitutions together.

	\begin{lemma}(Section V, \cite{sima2020optimal})\label{lemma:DS}
		For any binary sequence $\mathbf{x}=(x_1,x_2,\dots,x_n)\in\Sigma_2^n$,  there exists a function $h: \Sigma_2^n \rightarrow \Sigma_2^{[(t^2+1)(2t^2+1)+2t^2(t-1)]\log n+o(\log n)}$,  such that given $h(\mathbf{x})$ and $\mathbf{x}'$ obtained from $\mathbf{x}$ by $\tau_1$-deletions, $\tau_2$-insertions and $\tau_3$-substitutions satisfying $\tau_1+\tau_2+\tau_3=t$,
		one can uniquely recover $\mathbf{x}$.
	\end{lemma}

	For any $\mathbf{x}\in\Sigma_2^n$, define
	\begin{align}\label{eqn:N_T,t1}
		\scalebox{0.75}{$
		\mathcal{N}^{DI}_{2,(t_1,t_2)}(\mathbf{x})\triangleq\Big\{\mathbf{x}'\in\Sigma_2^n:\mathbf{x}'\ne \mathbf{x}, \mathcal{B}_{2,(t_1,t_2)}^{DI}(\mathbf{x}')\cap \mathcal{B}_{2,(t_1,t_2)}^{DI}(\mathbf{x})\ne \varnothing\Big\},
		$}
	\end{align}
where $t_1$ and $t_2$ are constants with respect to $n$.

		\begin{lemma}\label{lem:NDI}		
		For any $\mb{x}\in\Sigma_2^n$ and two nonnegative integers $t_1,t_2$, when $t_1>t_2\ge 2$, we have 
		\begin{align}\label{eqn:con2}
			\scalebox{0.8}{$
				\begin{aligned}
					\left|\mathcal{N}^{DI}_{2,(t_1,t_2)}(\mathbf{x})\right|\le& {n-2t_1+2\choose 2}{n-2t_1+2t_2+1\choose 2}\cdot 2^{2(t_1+t_2)-8}\\
					\le& n^42^{2(t_1+t_2)},
				\end{aligned}$}
		\end{align}
		and when $t_1>t_2=1$,
		\begin{align}\label{eqn:con21}
			\scalebox{0.8}{$
				\begin{aligned}
					\left|\mathcal{N}^{DI}_{2,(t_1,t_2)}(\mathbf{x})\right|\le& {n-2t_1+2\choose 2}{n-2t_1+2t_2+1\choose 2}\cdot 2^{2t_1-4}\\
					\le& n^42^{2t_1-4}.
				\end{aligned}$}
		\end{align}
		
	\end{lemma}
	\begin{proof}
		Each sequence in $\mathcal{N}^{DI}_{2,(t_1,t_2)}(\mathbf{x})$ can be obtained via the following  steps:
		\begin{itemize}
			\item Delete two substrings of length $t_1$ from $\mathbf{x}$,
			and then insert two substrings of length $t_2$ at the corresponding deletion positions to obtain a sequence $\mathbf{y}\in\Sigma_2^{n-2t_1+2t_2}$, where the first and last symbols of each inserted substring are different from the first and last symbols of the deleted substring at the same position, respectively.   
			There are at most ${n-2t_1+2\choose 2}2^{2t_2-4}$ and ${n-2t_1+2\choose 2}$ possibilities for $\mathbf{y}$ when $t_2\ge 2$ and $t_2=1$, respectively.
			
			\item Insert two substrings  of length $t_1$ and then delete two substrings of length $t_2$ from $\mathbf{y}$ following the insertion positions to obtain a sequence $\mathbf{x}'\in \Sigma_2^n$, where the first and last symbols of each inserted substring are different from the first and last symbols of the deleted substring at the same position, respectively.  
			There are at most ${n-2t_1+2t_2+1\choose 2}\cdot 2^{2t_1-4}$ possibilities for $\mathbf{x}'$ when $t_1\ge 2$.
		\end{itemize}

		The above argument shows that 
		\begin{align*}
			\scalebox{0.8}{$
				\begin{aligned}	\left|\mathcal{N}^{DI}_{2,(t_1,t_2)}(\mathbf{x})\right|\le&  {n-2t_1+2\choose 2}{n-2t_1+2t_2+1\choose 2}\cdot 2^{2t_1+2t_2-8}
				\end{aligned}
				$}
		\end{align*}
		for $t_1>t_2\ge 2$, and
		\begin{align*}
			\scalebox{0.8}{$
				\begin{aligned}	\left|\mathcal{N}^{DI}_{2,(t_1,t_2)}(\mathbf{x})\right|\le&  {n-2t_1+2\choose 2}{n-2t_1+2t_2+1\choose 2}\cdot 2^{2t_1-4}
				\end{aligned}
				$}
		\end{align*}
		for $t_1>t_2=1$.
		
		This completes the proof. 
		\end{proof}


	\begin{construction}\label{con2}
		For any $\mathbf{x}\in\Sigma_2^n$, let $h(\mathbf{x})$ be the function proposed in Lemma \ref{lemma:DS} for correcting $2t_1$-deletions and $2t_2$-insertions, i.e., $(\tau_1,\tau_2,\tau_3)=(2t_1,2t_2,0)$ and $t=2t_1+2t_2$, 
		and $\mathcal{N}^{DI}_{2,(t_1,t_2)}(\mathbf{x})$ be the set defined in \eqref{eqn:N_T,t1}. Denote 
		\begin{align}\label{eqn:Lhalpha1}
			f_2(\mathbf{x})=\Big(a(\mathbf{x}),h(\mathbf{x})\bmod a(\mathbf{x})\Big),
		\end{align}
		where $a(\mathbf{x})$ is a positive integer satisfying 
		\begin{align*}
			a(\mathbf{x})\le 2^{\log{\left|\mathcal{N}^{DI}_{2,(t_1,t_2)}(\mathbf{x})\right|+o(\log n)}}
		\end{align*}  and $a(\mathbf{x}) \notin\{j:j\mid (h(\mathbf{x})-h(\mathbf{x}'))\}$
		for any $\mathbf{x}'\in\mathcal{N}^{DI}_{2,(t_1,t_2)}(\mathbf{x})$.
		
		For a fixed parameter $f_2\in[0,(a(\mathbf{x}))^2)$, define a code by
		\begin{align*}
			\mathcal{C}_1\triangleq\Big\{\mathbf{x}\in\Sigma_2^n:f_2(\mathbf{x})=f_2\Big\}.
		\end{align*}
	\end{construction}

	\begin{theorem}\label{thm:Tts1}
	The code $\mathcal{C}_1$ generated in	Construction  \ref{con2} can correct two bursts of $(t_1,t_2)$-DI in a binary sequence of length $n$ with at most $8\log n +o(\log n)$ bits of redundancy.
	\end{theorem}
	\begin{proof}
		Since the function $h(\mathbf{x})$ employed in \eqref{eqn:Lhalpha1} can correct $2t_1$-deletions and $2t_2$-insertions, it can correct two bursts of $(t_1,t_2)$-DI.
		Viewing $h(\mathbf{x})$ as the binary representation of a nonnegative integer, by brute-force search, one can find a positive integer  $a(\mathbf{x})$ in time $$2^{\log{\left|\mathcal{N}^{DI}_{2,(t_1,t_2)}(\mathbf{x})\right|+o(\log n)}}= O(n^{4}).$$
		Since $h(\mathbf{x})\ne h(\mathbf{x}'')$, we have $f_2(\mathbf{x})\ne f_2(\mathbf{x}'')$ for $\mathbf{x}''\in\mathcal{N}^{DI}_{2,(t_1,t_2)}(\mathbf{x})$.
		Equivalently, the original sequence $\mathbf{x}$  can be uniquely recovered by $f_2(\mathbf{x})$ and $\mathbf{x}'\in \mathcal{B}_{2,(t_1,t_2)}^{DI}(\mathbf{x})$.

		From Lemma \ref{lem:NDI}, the redundancy of $f_2(\mathbf{x})$ in \eqref{eqn:Lhalpha1} is 
		\begin{align*}
			2\left|a(\mathbf{x})\right|\le & 2 \log\left|\mathcal{N}^{DI}_{2,(t_1,t_2)}(\mathbf{x})\right|+o(\log n)\\
			\le& 8\log n +o(\log n),
		\end{align*}
		completing the proof.
	\end{proof}
	
	 The following result can be obtained  by an argument similar to the one used in Construction \ref{con2} and Theorem \ref{thm:Tts1}.
	\begin{corollary}\label{coro}
		For any $\mathbf{x}\in\Sigma_2^n$, there exists a function $f_1:\Sigma_2^n\rightarrow\Sigma_2^{4\log n+o(\log n)}$ such that given $f_1(\mb{x})$ and $\mb{x}'\in\mathcal{B}_{(t_1,t_2)}^{DI}(\mb{x})$, one can recover the original sequence $\mb{x}$. In addition, the function $f_1(\mathbf{x})$ can be computed in $O(n^2)$ time.
	\end{corollary}

	
	While the functions derived from the syndrome compression technique are indeed capable of correcting errors in a sequence, applying this technique directly to binary sequences typically results in prohibitively high computational complexity, making their practical implementation challenging. Throughout this paper, for the syndrome-compression-based construction, computational complexity specifically refers to the theoretical complexity of the brute-force search required by the syndrome compression framework to determine the auxiliary modulus $a(\mb{x})$. It does not represent an overall implementation complexity of the encoder or decoder, which additionally depends on candidate generation, syndrome computation, verification, storage, and implementation-dependent optimizations.
	To address this issue, we construct ECCs capable of correcting two bursts of  $(t_1,t_2)$-DI by leveraging a matrix-based representation of the sequences in the next section.

As a basis for the next section, we note that the following conclusion can be obtained by an argument similar to that used in Construction \ref{con2}. 
	
		For any $\mathbf{x}\in\Sigma_2^n$, define
	\begin{align*}
		\scalebox{0.83}{$
			\mathcal{N}^{DS}_{2,(1,t)}(\mathbf{x})\triangleq\Big\{\mathbf{x}'\in\Sigma_2^n:\mathbf{x}'\ne \mathbf{x}, \mathcal{B}_{2,(1,t)}^{DS}(\mathbf{x}')\cap \mathcal{B}_{2,(1,t)}^{DS}(\mathbf{x})\ne \varnothing\Big\},
			$}
	\end{align*}
	where $t\ge 1$.
\begin{lemma}\label{lem:NDS}		
		For any $\mathbf{x}\in\Sigma_2^n$ and a positive integer $t\ge 1$, we have 
		\begin{align}\label{eqn:Nt1}
			\scalebox{0.9}{$
				\begin{aligned}
					\left|\mathcal{N}^{DS}_{2,(1,t)}(\mathbf{x})\right|\le& 4\cdot{n-2t\choose 2}{n-2t+1\choose 2}\Big(\sum_{i=0}^{t-1}{t-1\choose i}\Big)^4\\
					\le& n^42^{4t-4}.
				\end{aligned}$}
		\end{align}
	\end{lemma}	
	\begin{proof}
		Each sequence in $\mathcal{N}^{DS}_{2,(1,t)}(\mathbf{x})$ can be obtained via the following  steps:
		\begin{itemize}
			\item Delete two symbols from $\mathbf{x}$, and then substitute at most $t$ symbols following the deleted positions to obtain a sequence $\mathbf{y}\in \Sigma_2^{n-2}$, where the first substituted symbol is different from the deleted one at the same position.
			There are at most ${n-2t\choose 2}\left(\sum_{i=0}^{t-1}{t-1\choose i}\right)^2$ possibilities for $\mathbf{y}$.
			
			\item Insert two symbols into $\mathbf{y}$, and then substitute at most $t$ symbols following the inserted positions to obtain a sequence $\mathbf{x}'\in \Sigma_2^n$, where the first substituted symbol is different from the inserted one at the same position.  There are at most $4{n-2t+1\choose 2}\left(\sum_{i=0}^{t-1}{t-1\choose i}\right)^2$ possibilities for $\mathbf{x}'$.
		\end{itemize}
		Thus, we have 
		\begin{align*}
			\scalebox{0.95}{$
				\begin{aligned}
					\left|\mathcal{N}^{DS}_{2,(1,t)}(\mathbf{x})\right|\le&4\cdot {n-2t\choose 2}{n-2t+1\choose 2}\Big(\sum_{i=0}^{t-1}{t-1\choose i}\Big)^4.
				\end{aligned}$}
		\end{align*}
		This completes the proof.
	\end{proof}

Therefore, the following lemma holds.
		\begin{lemma}\label{thm:Tt}
		For any binary sequence $\mathbf{x}\in\Sigma_2^n$, there exists a function $f:\Sigma_2^n\rightarrow \Sigma_2^{8\log n+o(\log n)}$, such that given $\mathbf{x}'\in \mathcal{B}_{2,(1,t)}^{DS}(\mathbf{x})$ and $f(\mathbf{x})$, one can  uniquely recover $\mathbf{x}$.
	\end{lemma}

	\section{Two-Burst $(t_1,t_2)$-DI ECCs with Reduced Complexity}\label{sec:T=2}
	
	In this section, we present constructions of binary ECCs capable of correcting two bursts of $(t_1,t_2)$-DI, where $t_1$ and $t_2$ are positive integers that do not depend on $n$ with $t_1 \geq t_2$.  

	\subsection{Two-Burst $(t,t)$-DI ECCs}
	In this subsection, we first construct binary codes correcting two bursts of $(t,t)$-DI, i.e., $t_1=t_2=t$. 
	In fact, two bursts of  $(t,t)$-DI in a binary sequence are equivalent to two bursts of $t$-substitutions, in which the first and the last symbols are necessarily substituted, which can be converted to two bursts of $(2,2)$-DI in the corresponding $2^{t-1}$-ary sequence following the results in \cite{sun2024asymptotically}. 

	
	For any binary sequence $\mathbf{x}=(x_1,x_2,\dots,x_n)\in\Sigma_2^n$, assume that $(t-1)\mid n$; otherwise, we append zeros at the end of each sequence so that its length is the smallest multiple of $t-1$ that is not less than $n$.
	Then, when $t\ge 2$, the binary sequence $\mathbf{x}\in\Sigma_2^n$ can be represented by a $(t-1)\times (n/ (t-1))$ binary matrix as follows
	\begin{align*}
		X_{t-1}=\left(\begin{array}{cccccc}
			x_1 & x_{t} &\cdots & x_{n-t+2}\\
			x_2 & x_{t+1} &\cdots & x_{n-t+3}\\
			\vdots & \vdots &\ddots &\vdots\\
			x_{t-1}& x_{2t-2} &\cdots&  x_{n}
		\end{array}\right).
	\end{align*}

	Denote a $2^{t-1}$-ary vector corresponding to $X_{t-1}$ by
	\begin{align}\label{eqn:hat{x}}
		\tilde{\mathbf{x}}=(\tilde{x}_1,\tilde{x}_2,\dots,\tilde{x}_{n\over {t-1}})\in\Sigma_{2^{t-1}}^{n\over {t-1}},
	\end{align} 
	where 
	$$\tilde{x}_i=\sum_{j=1}^{t-1}2^{t-1-j}\cdot x_{(i-1)(t-1)+j},\quad i\in\left[{n\over {t-1}}\right].$$
	\begin{lemma}[Theorem 3, \cite{sun2024asymptotically}]\label{lem:(t,t)}
		Let $\mathbf{x}$ be a binary sequence of length $n$ and $\tilde{\mathbf{x}}$ be the corresponding $2^{t-1}$-ary vector defined in \eqref{eqn:hat{x}}, where $t\ge2$ is a positive integer. If a burst of $(t, t)$-DI occurs in $\mathbf{x}$, then there is a burst of $(2,2)$-DI in $\tilde{\mathbf{x}}$.
	\end{lemma}

	\begin{theorem}\label{thm:t1=t2}
		For $t\ge 2$ and $\mathbf{u}\in\mathbb{F}_q^{n/(t-1)}$, define
		\begin{align*}
			\mathcal{C}^{\mathbf{u}}=\Big\{\mathbf{x}\in\Sigma_2^n:\tilde{\mathbf{x}}\in\{\mathbf{y}+\mathbf{u}:\mathbf{y}\in\mathcal{C}_1\}\Big\},
		\end{align*}
		where $\mathcal{C}_1$ is an $[n/(t-1),n/(t-1)-8,d_{\mathrm{min}}=9]_q$ Reed-Solomon (RS) code with $q\ge n/(t-1)$ being a prime power.
		Then, $\mathcal{C}^{\mb{u}}$ is an ECC capable of correcting two bursts of $(t,t)$-DI with at most  $8\log n+O(1)$ bits of redundancy.
	\end{theorem}
\begin{proof}
	According to Lemma \ref{lem:(t,t)}, two bursts of $(t,t)$-DI in  the original sequence $\mathbf{x}\in\Sigma_2^n$ induce two bursts of $(2,2)$-DI in the transformed sequence $\tilde{\mathbf{x}}$.
	These errors manifest as two bursts of two-symbol substitutions in $\tilde{\mathbf{x}}$, which together amount to a total of four symbol substitutions. Such error patterns can be corrected since $\tilde{\mathbf{x}}\in\{\mathbf{y}+\mathbf{u}:\mathbf{y}\in\mathcal{C}_1\}$ and $d_{\mathrm{min}}(\mathcal{C}_1)=9$.

	Since $\mathcal{C}_{1}$ is a subspace of $\mathbb{F}_q^{n/(t-1)}$ with dimension $n/(t-1)-8$, its $q^{8}$ cosets form a partition of $\mathbb{F}^{n/(t-1)}_q$. That is, there exist some vectors $\mathbf{u}_i\in \mathbb{F}_q^{n/(t-1)}$ such that 
	$\mathcal{C}^{\mathbf{u}_i} \triangleq \{\mathbf{y}+\mathbf{u}_i:\mathbf{y}\in\mathcal{C}_1\}$,
	and the cosets $\mathcal{C}^{\mathbf{u}_i}$ constitute a partition of $\mathbb{F}_{q}^{n/(t-1)}$, where $i\in [q^{8}]$.
	Note that each coset $\mathcal{C}^{\mathbf{u}_i}$ has the same error-correcting capability as $\mathcal{C}_1$ since $d_{\mathrm{min}}(\mathcal{C}^{\mathbf{u}_i})=d_{\mathrm{min}}(\mathcal{C}_{1})$.
	By the pigeonhole principle, there exists an $i\in[q^{8}]$ such that 
	\begin{align*}
		|\mathcal{C}^{\mathbf{u}_i}|\ge {|\cup_{i\in [q^8]}\mathcal{C}^{\mathbf{u}_i}|\over q^{8}}.	
	\end{align*}
	Consequently, 
	there exists a choice of $\mathbf{u}=\mathbf{u}_i$ for which the corresponding coset contributes  at most $8\log n+O(1)$ bits of redundancy.	
\end{proof}

	\begin{remark}
		When $t=1$, two bursts of $(1,1)$-DI in $\mathbf{x}\in\Sigma_2^n$ cause two substitutions in $\mathbf{x}$, and therefore, applying a coset of an $[n,n-4,d_{\mathrm{min}}=5]_q$ RS code with a prime power $q\ge n$ is sufficient, which results in at most $4\log n+O(1)$ bits of redundancy.
	\end{remark}

	\subsection{Two-Burst $(t_1,t_2)$-DI ECCs}
	In this subsection, we consider the case of $t_1>t_2$ and construct ECCs capable of correcting two bursts of $(t_1,t_2)$-DI errors.
	By representing a sequence in matrix form, we can demonstrate that two bursts of $(t_1,t_2)$-DI  cause two bursts of $(1, t'-1)$-DS in each row of the matrix, where $t'=\lceil t_1/(t_1-t_2)\rceil$. Then, by locating the positions of the two bursts in the first row, the error positions of the original sequence can be approximately identified. Finally, two bursts of $(t_1,t_2)$-DI in the original sequence  can be corrected following the function proposed in Construction \ref{con2} with improved computational efficiency compared with the codes obtained by using the syndrome compression technique directly.
	
	For any binary sequence $\mathbf{x}=(x_1,x_2,\dots,x_n)\in\Sigma_2^n$, assume that $(t_1-t_2)\mid n$; otherwise, we append zeros at the end of each sequence so that its length is the smallest multiple of $t_1-t_2$ that is not less than $n$.
	Then, the binary sequence $\mathbf{x}\in\Sigma_2^n$ can be represented by a $(t_1-t_2)\times (n/(t_1-t_2))$ binary matrix as follows
		\begin{align}\label{eqn:matrix}
			\scalebox{0.8}{$
				\begin{aligned}
		X_{t_1-t_2}&=\left(\begin{array}{cccccc}
			X[1] \\
			X[2]\\
			\vdots\\
			X[t_1-t_2]
		\end{array}\right)=\left(\begin{array}{cccccc}
			x_1 &\cdots & x_{n-t_1+t_2+1}\\
			x_2 & \cdots & x_{n-t_1+t_2+2}\\
			\vdots  &\ddots &\vdots\\
			x_{t_1-t_2} &\cdots&  x_{n}
		\end{array}\right),
		\end{aligned}	
$}
\end{align}	
	where $X[i],i\in[t_1-t_2],$ refers to the $i$-th row of $X_{t_1-t_2}$.
	
	Let $t'=\lceil t_1/(t_1-t_2)\rceil$, and
	\begin{align*}
		\mathcal{M}=\left\{m\in [n]:m=a(t_1-t_2)+1,a\in\left[0,{n\over t_1-t_2}\right)\right\}.
	\end{align*}
	The following fact can be obtained by analyzing the matrix representation of $\mathbf{x}$ in \eqref{eqn:matrix}.
	
	\begin{observation}\label{fact:location}
		For any binary sequence $\mathbf{x}\in\Sigma_2^n$ and $i\in[n-t_1+1]$,
		suppose that one burst of $(t_1,t_2)$-DI occurs at the $i$-th component of $\mathbf{x}$. Then,
		one burst of $(1,t'-1)$-DS occurs at the $b_1$-th  component of $X[1]$ when $i\in\mathcal{M}$, where $b_1=\lceil i/(t_1-t_2) \rceil$, or the $b_2$-th component of $X[1]$ when $i\in [n]\backslash\mathcal{M}$, where  $b_2\in[\lceil i/(t_1-t_2)\rceil+1,\lceil i/(t_1-t_2)\rceil+t')$.
		Moreover, the starting  position of the errors in $X[j]$, $j\in[2,t_1-t_2]$, is at most $t'$ columns before the starting  position of the errors in $X[1]$.
	\end{observation}
	
	\begin{example}
		For $n=18,t_1=10,t_2=7$, we have $t'=4$. Assume that $$\mathbf{x}=(0,1,1,0,0,0,0,0,1,1,0,1,1,0,1,0,1,1).$$
		If $(0,0,0,0,0,1,1,0,1,1)$ is deleted and $(1,0,1,0,1,1,0)$  is inserted at the fourth component of $\mathbf{x}$, that is,
		\begin{align*}
			X_{3}=\left(\begin{array}{ccccccccccc}
				0 & \underline{0} & \underline{0} & \underline{1} & \underline{1} & 0\\
				1 & \underline{0} & \underline{0} & \underline{0} & 0 & 1 \\
				1 & \underline{0} & \underline{1} & \underline{1} & 1 & 1
			\end{array}\right)\rightarrow
			\left(\begin{array}{ccccccccccc}
				0 & \bar{1} & \bar{0} & \bar{0} & 0\\
				1 & \bar{0} & \bar{1} & 0 & 1 \\
				1 & \bar{1} & \bar{1} & 1 & 1
			\end{array}\right),
		\end{align*}
			where $\underline{i}$ is the deleted bit and $\bar{i}$ is the inserted bit in $\mathbf{x}$, $i\in\{0,1\}$,	then there is one burst of $(1,3)$-DS at the second component of $X[1]$ and the errors in $X[2]$ and 
		$X[3]$ begin at the third element of the second row and the second component of the third row, respectively.

		If $(0,0,0,0,1,1,0,1,1,0)$ is deleted and $(1,1,1,0,0,0,1)$, $(1,1,0,0,0,0,1)$ or $(1,0,0,1,1,1,1)$ is inserted as follows:
\begin{align}
	\scalebox{0.9}{$
		X_{3}=
		\begin{pmatrix}
			0 & 0 & \underline{0} & \underline{1} & \underline{1} & 0\\
			1 & \underline{0} & \underline{0} & \underline{0} & \underline{0} & 1\\
			1 & \underline{0} & \underline{1} & \underline{1} & 1 & 1
		\end{pmatrix}
		$}
	&\rightarrow
	\scalebox{0.9}{$
		\begin{pmatrix}
			0 & 0 & \bar{1} & \bar{0} & 0\\
			1 & \bar{1} & \bar{0} & \bar{1} & 1\\
			1 & \bar{1} & \bar{0} & 1 & 1
		\end{pmatrix}
		$}
	\label{DS-case1}
	\\
	&\rightarrow
	\scalebox{0.9}{$
		\begin{pmatrix}
			0 & 0 & \bar{0} & \bar{0} & 0\\
			1 & \bar{1} & \bar{0} & \bar{1} & 1\\
			1 & \bar{1} & \bar{0} & 1 & 1
		\end{pmatrix}
		$}
	\label{DS-case2}
	\\
	&\rightarrow
	\scalebox{0.9}{$
		\begin{pmatrix}
			0 & 0 & \bar{0} & \bar{1} & 0\\
			1 & \bar{1} & \bar{1} & \bar{1} & 1\\
			1 & \bar{0} & \bar{1} & 1 & 1
		\end{pmatrix},
		$}
	\label{DS-case3}
\end{align}
	then one burst of $(1,2)$-DS, one burst of $(1,1)$-DS and $1$-deletion
		occur at the third, fourth, and fifth components of $X[1]$ in \eqref{DS-case1}, \eqref{DS-case2} and \eqref{DS-case3}, respectively, and the errors in $X[2]$ and $X[3]$ start  at most three columns before the starting  position of the error in $X[1]$.
	\end{example}

In this section, we focus on binary sequences whose representation under $X_{t_1-t_2}$ has a $d$-regular first row. The following lemma presents an explicit encoding function that transforms arbitrary binary sequences into ones whose first row of the matrix representation is $d$-regular.

	\begin{lemma}\label{lem:regular}
		For any binary sequence $\mathbf{x}\in\Sigma_2^n$, there exists a one-to-one mapping $\mathcal{E}:\Sigma_2^n\rightarrow \Sigma_2^{n+1}$,  such that
		for $\mathbf{x}'=\mathcal{E}(\mathbf{x})$, if $X'_{t_1-t_2}$ is the matrix representation of  $\mathbf{x}'$,  then $X'[1]$ is a $d$-regular sequence. 
	\end{lemma}
	\begin{proof}
		Following  \eqref{eqn:matrix}, 
		we apply the encoding function $\mathrm{RegEnc}$ in  Lemma \ref{regular} to $X[1]$ such that 
		\begin{align*}
			\mathrm{RegEnc}(X[1])=(u_1,u_2,\cdots,u_{n/(t_1-t_2)},u_{n/(t_1-t_2)+1})
		\end{align*} 
		is a $d$-regular sequence. Then, let
		\begin{align*}
			\scalebox{0.85}{$
				\begin{aligned}
					X'_{t_1-t_2}=\left(\begin{array}{cccccc}
						u_1 &\cdots & u_{n/(t_1-t_2)} & u_{n/(t_1-t_2)+1} \\
						x_2 & \cdots & x_{n-t_1+t_2+2} & 0\\
						\vdots  &\ddots &\vdots &\vdots\\
						x_{t_1-t_2} &\cdots&  x_{n} & 0
					\end{array}\right),
				\end{aligned}$}
		\end{align*}
		which corresponds to a sequence $\mathbf{x}'=(x'_1,x'_2,\dots,x'_{n+1})$ with $x'_i=x_i$ for
		$$i\in[n]\backslash\{(j-1)(t_1-t_2)+1:j\in[n/(t_1-t_2)]\},$$
		$x'_{(j-1)(t_1-t_2)+1}=u_j$ for $j\in[n/(t_1-t_2)]$, and $x'_{n+1}=u_{n/(t_1-t_2)+1}$.

		Letting $\mathbf{x}'=\mathcal{E}(\mathbf{x})$, the existence of such a function is guaranteed.
		Since $\mathrm{RegEnc}$ is a one-to-one mapping and all the remaining symbols are copied unchanged, the mapping $\mathcal{E}$ is injective. Moreover, the function $\mathcal{E}$ increases the sequence length by only one bit. Finally, as $\mathrm{RegEnc}$ can be computed in near-linear time, the same holds for $\mathcal{E}$.
\end{proof}


	In the following, we consider two bursts of $(t_1,t_2)$-DI in a binary sequence $\mathbf{x}\in\Sigma_2^n$. Recall that the two bursts under consideration in this subsection are mutually non-overlapping.
	To proceed, we need the following lemma.

	\begin{lemma}\label{lem:2(1,t')-DS}
		For a $d$-regular sequence $\mathbf{x}\in\Sigma_2^n$, if $\mathbf{x}'$ is obtained from $\mathbf{x}$ by two bursts of $(1,t-1)$-DS, then one of the following conclusions holds:
		\begin{itemize}
			\item[a)] There are  two distinct intervals $\mathcal{I}_{1},\mathcal{I}_2\subseteq[n]$ with $|\mathcal{I}_1|,|\mathcal{I}_2|\le d\log n$,  such that one burst of $(1,t-1)$-DS occurs in $\mathbf{x}_{\mathcal{I}_1}$ and the other occurs in $\mathbf{x}_{\mathcal{I}_2}$, respectively.  
			\item[b)] There is an interval $\mathcal{I}\subseteq[n]$ with $|\mathcal{I}|\le 3d\log n+2t$, such that  two bursts of $(1,t-1)$-DS occur in $\mathbf{x}_{\mathcal{I}}$. 
		\end{itemize}
	\end{lemma}
	\begin{proof}
		Assume that $\mathbf{x}'$ is obtained from $\mathbf{x}$ by two bursts of $(1,t-1)$-DS at the $i_1$-th and $i_2$-th components, where $1\le i_1\le i_2-t\le n-2t+1$.
		Suppose that there also exist two indices $j_1,j_2$ such that $\mathbf{x}'$ can be obtained from $\mathbf{x}$ by two bursts of $(1,t-1)$-DS at the $j_1$-th and $j_2$-th components, where $1\le j_1\le j_2-t\le n-2t+1$. 
		In what follows, we analyze the cases where the four burst positions are either disjoint or overlapping. 
		For the disjoint case, we distinguish the following three cases due to symmetry.
		
\textbf{Case i) $i_1+t\le j_1\le i_2-t\le j_2-2t$.} In this case, we have
		\begin{align*}
			\mathbf{x}'=&x_1\cdots x_{i_1-1}\overline{x_{i_1+1}\cdots x_{i_1+t-1}}~x_{i_1+t}~~\cdots ~~x_{j_1}~x_{j_1+1}\cdots\\
			& x_{j_1+t-1}x_{j_1+t}\cdots x_{i_2-1}\overline{x_{i_2+1}\cdots x_{i_2+t-1}}~x_{i_2+t}\cdots~ x_{j_2}~~\\
			&x_{j_2+1}\cdots x_{j_2+t-1}x_{j_2+t}\cdots x_n \\
			=&x_1\cdots x_{i_1-1}~~x_{i_1}~\cdots x_{i_1+t-2}x_{i_1+t-1}\cdots x_{j_1-1}\overline{x_{j_1+1}\cdots}\\
			&\overline{x_{j_1+t-1}}x_{j_1+t}\cdots x_{i_2-1}~~x_{i_2}~\cdots x_{i_2+t-2}x_{i_2+t-1}\cdots x_{j_2-1}\\
			&\overline{x_{j_2+1}\cdots x_{j_2+t-1}}x_{j_2+t}\cdots x_n,
		\end{align*}
		where $\overline{x_{i+1}\cdots x_{i+t-1}}$ means that at most $t-1$ substitutions occur in $(x_{i+1}\cdots x_{i+t-1}),i\in\{i_1,i_2,j_1,j_2\}$. 
		
		Viewing $\mathbf{x}'$, it is impossible that $\overline{x_{i_1+1}\cdots x_{i_1+t-1}}=x_{i_1}\cdots x_{i_1+t-2}$ and $\overline{x_{i_2+1}\cdots x_{i_2+t-1}}=x_{i_2}\cdots x_{i_2+t-2}$ since $\overline{x}_{i_1+1}\ne x_{i_1}$ and $\overline{x}_{i_2+1}\ne x_{i_2}$ according to Definition \ref{def:DS}. This indicates that two bursts of $(1,t-1)$-DS in $\mathbf{x}$ reduce to two deletions in this case, i.e.,
		\begin{align*}
			\mathbf{x}'=&x_1\cdots x_{i_1-1}x_{i_1+1}\cdots ~~x_{j_1}~x_{j_1+1}\cdots  x_{i_2-1}x_{i_2+1}\cdots~\\
			& ~x_{j_2}~~x_{j_2+1}\cdots x_n \\
			=&x_1\cdots x_{i_1-1}~~x_{i_1}~\cdots x_{j_1-1}x_{j_1+1}\cdots x_{i_2-1}~~x_{i_2}~\cdots \\
			& x_{j_2-1}x_{j_2+1}\cdots x_n.
		\end{align*}

		By comparing the symbols of the corresponding positions, we obtain
		\begin{equation}
			\begin{aligned}\label{eqn:i1j1}
				x_{i_1}=x_{i_1+1}=\cdots= x_{j_1-1}=x_{j_1},\\
				x_{i_2}=x_{i_2+1}=\cdots= x_{j_2-1}=x_{j_2}.
			\end{aligned}
		\end{equation}

		Following \eqref{eqn:i1j1}, there exist two runs $\mb{x}_{[a',b']}$ and $\mb{x}_{[a'',b'']}$ such that   $[i_1,j_1]\subseteq[a',b']$ and $[i_2,j_2]\subseteq[a'',b'']$.
		Denoting $\mathcal{I}_1=[a',b']$ and $\mathcal{I}_2=[a'',b'']$,
		we have $|\mathcal{I}_1|,|\mathcal{I}_2|\le d\log n$ by Observation \ref{observation}. Thus, conclusion a) holds.

	\textbf{ Case ii) $i_1+t\le j_1\le j_2-t\le i_2-2t$.} In this case, we have
		\begin{align*}
			\mathbf{x}'=&x_1\cdots x_{i_1-1}\overline{x_{i_1+1}\cdots x_{i_1+t-1}}~x_{i_1+t}~~\cdots ~~x_{j_1}~x_{j_1+1}\cdots\\
			& x_{j_1+t-1}x_{j_1+t}\cdots x_{j_2-1}~x_{j_2}~\cdots x_{j_2+t-2}x_{j_2+t-1}\cdots x_{i_2-1}\\
			&\overline{x_{i_2+1}\cdots x_{i_2+t-1}}x_{i_2+t}\cdots x_n\\
			=&x_1\cdots x_{i_1-1}~~x_{i_1}~\cdots x_{i_1+t-2}x_{i_1+t-1}\cdots x_{j_1-1}\overline{x_{j_1+1}\cdots}\\
			&\overline{x_{j_1+t-1}}x_{j_1+t}\cdots x_{j_2-1}\overline{x_{j_2+1}\cdots x_{j_2+t-1}}~~x_{j_2+t}\cdots ~~ x_{i_2}~\\
			&x_{i_2+1}\cdots x_{i_2+t-1}x_{i_2+t}\cdots x_n.
		\end{align*}
	
		 Viewing $\mathbf{x}'$, two bursts of $(1,t-1)$-DS in $\mathbf{x}$ reduce to two deletions by an argument similar to that used in  Case i).
			Let $\mathcal{I}_1=[a',b']$ and $\mathcal{I}_2=[a'',b'']$ such that   $[i_1,j_1]\subseteq[a',b']$ and $[j_2,i_2]\subseteq[a'',b'']$. Then, we have $|\mathcal{I}_1|,|\mathcal{I}_2|\le d\log n$. Thus, conclusion a) holds.
		
\textbf{Case iii) $i_1+t\le i_2\le j_1-t+2\le j_2-2t+2$.} In this case,
		\begin{align*}
			\mathbf{x}'=&x_1\cdots x_{i_1-1}\overline{x_{i_1+1}\cdots x_{i_1+t-1}}~x_{i_1+t}~~\cdots  x_{i_2-1}\overline{x_{i_2+1}\cdots}\\
			&\overline{x_{i_2+t-1}}~x_{i_2+t}~\cdots
			x_{j_1+1}x_{j_1+2}\cdots ~~x_{j_1+t}~x_{j_1+t+1}\cdots~~\\
			& x_{j_2}~x_{j_2+1}\cdots x_{j_2+t-1}x_{j_2+t}\cdots x_n\\
			=&x_1\cdots x_{i_1-1}~~x_{i_1}~\cdots x_{i_1+t-2}x_{i_1+t-1}\cdots x_{i_2-2}x_{i_2-1}\cdots\\
			& x_{i_2+t-3}x_{i_2+t-2}\cdots x_{j_1-1}\overline{x_{j_1+1}\cdots x_{j_1+t-1}}~~x_{j_1+t}~\cdots\\
			&  x_{j_2-1}\overline{x_{j_2+1}\cdots x_{j_2+t-1}}x_{j_2+t}\cdots x_n.
		\end{align*}
	
	 That is, two bursts of $(1,t-1)$-DS in $\mathbf{x}$ reduce to the following form,
		\begin{align*}
		\mathbf{x}'=&x_1\cdots x_{i_1-1}x_{i_1+1}\cdots  x_{i_2-1}\overline{x_{i_2+1}\cdots x_{i_2+t-1}}~~x_{i_2+t}~\cdots\\
		&
		x_{j_1+1}x_{j_1+2}\cdots ~~x_{j_1+t}~x_{j_1+t+1}\cdots~~x_{j_2}~x_{j_2+1}\cdots x_n\\
		=&x_1\cdots x_{i_1-1}~~x_{i_1}~\cdots x_{i_2-2}x_{i_2-1}\cdots x_{i_2+t-3}x_{i_2+t-2}\cdots\\
		& x_{j_1-1}\overline{x_{j_1+1}\cdots x_{j_1+t-1}}~~x_{j_1+t}~\cdots  x_{j_2-1} x_{j_2+1}\cdots x_n.
	\end{align*}

		By comparing the symbols of the corresponding positions, we obtain
		\begin{equation}
			\begin{aligned}\label{eqn:i1i2}
				x_{i_1}=x_{i_1+1}=\cdots= x_{i_2-2}=x_{i_2-1},\\
				x_{j_1+t}=x_{j_1+t+1}=\cdots= x_{j_2-1}=x_{j_2}.
			\end{aligned}
		\end{equation}
		Meanwhile, if $j_1-i_2-t+1$ is even, we have
		\begin{equation}\label{eqn:i2j1}
			\begin{aligned}
				x_{i_2+t-2}=x_{i_2+t}=\cdots=x_{j_1-1}=x_{j_1+1},\\
				x_{i_2+t-1}=x_{i_2+t+1}=\cdots=x_{j_1-2}=x_{j_1}.
			\end{aligned}
		\end{equation}
		If $j_1-i_2-t+1$ is odd, we have
		\begin{equation}\label{eqn:i2j11}
			\begin{aligned}
				x_{i_2+t-2}&=x_{i_2+t}=\cdots=x_{j_1-2}=x_{j_1},\\
				x_{i_2+t-1}&=x_{i_2+t+1}=\cdots=x_{j_1-1}=x_{j_1+1}.
			\end{aligned}
		\end{equation}

		Regardless of whether $j_1-i_2-t+1$ is even or odd,  there exists a run $\mb{x}_{[a,b]}$ such that $[i_2+t-2,j_1+1]\subseteq[a,b]$ if $x_{j_1}=x_{j_1+1}$, otherwise there exists an alternating substring.
		In addition, there exist two runs $\mb{x}_{[a',b']}$ and $\mb{x}_{[a'',b'']}$ such that   $[i_1,i_2-1]\subseteq[a',b']$ and $[j_1+t,j_2]\subseteq[a'',b'']$.
		Denoting $\mathcal{I}=[a',b'']$, we have $|\mathcal{I}|\le  3d\log n+2t$
		due to Equations \eqref{eqn:i1i2}-\eqref{eqn:i2j11} and Observation \ref{observation}.
		Consequently, conclusion b) holds.

	Furthermore, for scenarios involving overlapping bursts, the underlying intuition is illustrated in Fig. \ref{fig:cases}.
		Following the same approach as the previous case i),  two bursts of $(1,t-1)$-DS errors in $\mathbf{x}$ in Fig. \ref{fig:cases}-i) and ii) reduce to two deletions, which can be handled correspondingly.
	Regarding Fig. \ref{fig:cases}-iii), we can observe that there exist two runs $\mb{x}_{[a',b']}$ and $\mb{x}_{[a'',b'']}$ such that   $[i_1,i_2-1]\subseteq[a',b']$ and $[j_1+t,j_2]\subseteq[a'',b'']$.
	  Hence, denoting $\mathcal{I} = [a', b'']$, conclusion b) follows.
		Since the details are similar to those of the non-overlapping cases, they are omitted for brevity.

			\begin{figure}[t]
			\centering
			\hspace*{-5cm}\begin{subfigure}{\textwidth}
				\centering
				\begin{tikzpicture}[scale=0.5, transform shape]	
					
					\draw[-] (0,0) -- (1,0);
					\node at (1,0.3) {$i_1-1$};	\fill (1,0) circle (1pt);
					\node at (2.3,0){$\bar{x}_{i_1+1},\dots,\bar{x}_{i_1+t}$};
					\draw[-] (3.6,0) -- (7,0);
					\node at (3.8,0.3) {$i_1+t+1$};\fill (3.6,0) circle (1pt);
					\node at (7,0.3) {$i_2-1$};\fill (7,0) circle (1pt);
					\node at (8.3,0){$\bar{x}_{i_2+1},\dots,\bar{x}_{i_2+t}$};
					\node at (10.0,0.3) {$i_2+t+1$};\fill (9.6,0) circle (1pt);
					\draw[-] (9.6,0) -- (15,0);
					
					\draw[-] (0,-1) -- (1.5,-1);
					\node at (1.5,-0.7) {$j_1-1$};\fill (1.5,-1) circle (1pt);
					\draw[-] (4.1,-1) -- (8.3,-1);
					\node at (2.8,-1){$\bar{x}_{j_1+1},\dots,\bar{x}_{j_1+t}$};
					\node at (4.1,-0.7) {$j_1+t+1$};\fill (4.1,-1) circle (1pt);
					\node at (7,-0.7) {$i_2-1$};\fill (7,-1) circle (1pt);
					\node at (8.3,-0.7) {$j_2-1$};\fill (8.3,-1) circle (1pt);
					\node at (9.6,-1){$\bar{x}_{j_2+1},\dots,\bar{x}_{j_2+t}$};
					\node at (10.9,-0.7) {$j_2+t+1$};\fill (10.9,-1) circle (1pt);
					\draw[-] (10.9,-1) -- (15,-1);
				\end{tikzpicture}
				\caption*{i) $i_1\le j_1\le i_2\le j_2~(j_1< i_1+t,j_2< i_2+t)$}
			\end{subfigure}
			\par
			\hspace*{-5cm}\begin{subfigure}{\textwidth}
				\centering
				\begin{tikzpicture}[scale=0.5, transform shape]	
					
					\draw[-] (0,0) -- (1,0);
					\node at (1,0.3) {$i_1-1$};	\fill (1,0) circle (1pt);
					\node at (2.3,0){$\bar{x}_{i_1+1},\dots,\bar{x}_{i_1+t}$};
					\draw[-] (3.6,0) -- (8.3,0);
					\node at (3.8,0.3) {$i_1+t+1$};\fill (3.6,0) circle (1pt);
					\node at (8.3,0.3) {$i_2-1$};\fill (8.3,0) circle (1pt);
					\node at (9.6,0){$\bar{x}_{i_2+1},\dots,\bar{x}_{i_2+t}$};
					\node at (10.9,0.3) {$i_2+t+1$};\fill (10.9,0) circle (1pt);
					\draw[-] (10.9,0) -- (15,0);
					
					\draw[-] (0,-1) -- (1.5,-1);
					\node at (1.5,-0.7) {$j_1-1$};\fill (1.5,-1) circle (1pt);
					\draw[-] (4.1,-1) -- (7,-1);
					\node at (2.8,-1){$\bar{x}_{j_1+1},\dots,\bar{x}_{j_1+t}$};
					\node at (4.1,-0.7) {$j_1+t+1$};\fill (4.1,-1) circle (1pt);
					\node at (7,-0.7) {$j_2-1$};\fill (7,-1) circle (1pt);
					\node at (8.3,-1){$\bar{x}_{j_2+1},\dots,\bar{x}_{j_2+t}$};
					\node at (10.0,-0.7) {$j_2+t+1$};\fill (9.6,-1) circle (1pt);
					\draw[-] (9.6,-1) -- (15,-1);
				\end{tikzpicture}
				\caption*{ii) $i_1\le j_1\le j_2\le i_2~(j_1< i_1+t,j_2>i_2-t)$}
			\end{subfigure}
			\par
			\hspace*{-5cm}\begin{subfigure}{\textwidth}
				\centering
				\begin{tikzpicture}[scale=0.5, transform shape]	
					
					\draw[-] (0,0) -- (1,0);
					\node at (1,0.3) {$i_1-1$};	\fill (1,0) circle (1pt);
					\node at (2.3,0){$\bar{x}_{i_1+1},\dots,\bar{x}_{i_1+t}$};
					\draw[-] (3.6,0) -- (7,0);
					\node at (3.8,0.3) {$i_1+t+1$};\fill (3.6,0) circle (1pt);
					\node at (7,0.3) {$i_2-1$};\fill (7,0) circle (1pt);
					\node at (8.3,0){$\bar{x}_{i_2+1},\dots,\bar{x}_{i_2+t}$};
					\node at (10.0,0.3) {$i_2+t+1$};\fill (9.6,0) circle (1pt);
					\draw[-] (9.6,0) -- (15,0);
					
					\draw[-] (0,-1) -- (7.5,-1);
					\node at (7.6,-0.7) {$j_1-1$};\fill (7.5,-1) circle (1pt);
					\draw[-] (10.1,-1) -- (11.0,-1);
					\node at (8.8,-1){$\bar{x}_{j_1+1},\dots,\bar{x}_{j_1+t}$};
					\node at (9.7,-0.7) {$j_1+t+1$};\fill (10.1,-1) circle (1pt);
					\node at (6.9,-0.7) {$i_2$};\fill (7,-1) circle (1pt);
					\node at (11.0,-0.7) {$j_2-1$};\fill (11,-1) circle (1pt);
					\node at (12.3,-1){$\bar{x}_{j_2+1},\dots,\bar{x}_{j_2+t}$};
					\node at (13.8,-0.7) {$j_2+t+1$};\fill (13.6,-1) circle (1pt);
					\draw[-] (13.6,-1) -- (15,-1);
				\end{tikzpicture}
				\caption*{iii) $i_1\le i_2 \le j_1\le j_2~(j_1 < i_2+t)$}
			\end{subfigure}
			\caption{Three different cases with intersections.}	
			\label{fig:cases}
		\end{figure}
		This completes the proof.
	\end{proof}
	
	
	
	The following conclusion is immediate from Lemma \ref{lem:2(1,t')-DS}.
	\begin{lemma}\label{thm:2bursts}
		For a binary sequence $\mathbf{x}\in\Sigma_2^n$ and two positive integers $t_1,t_2$ with $t_1>t_2$, let $X_{t_1-t_2}$ be the matrix representation of $\mathbf{x}$ following \eqref{eqn:matrix} and $f(X[1])$ be the function proposed in Lemma \ref{thm:Tt}. Let $t'=\lceil t_1/(t_1-t_2)\rceil$.
		Suppose that $X[1]$ is a $d$-regular sequence. Then if $\mathbf{x}'$ is obtained from  $\mathbf{x}$ by	two bursts of $(t_1,t_2)$-DI, one of the following conclusions holds:
		\begin{itemize}
			\item [a)] One can find two intervals $\mathcal{J}_1,\mathcal{J}_2\subseteq[n]$ with  $|\mathcal{J}_1|,|\mathcal{J}_2|\le (d\log (n/(t_1-t_2))+2t'+1)(t_1-t_2)$ such that 
			one burst of $(t_1,t_2)$-DI occurs in $\mathbf{x}_{\mathcal{J}_1}$ and the other occurs in $\mathbf{x}_{\mathcal{J}_2}$, respectively.
			\item [b)] One can find an interval $\mathcal{J}\subseteq[n]$ with $|\mathcal{J}|\le (3d\log {(n/ (t_1-t_2))}+4t'+1)(t_1-t_2)$ such that 
			 two bursts of $(t_1,t_2)$-DI occur in $\mathbf{x}_{\mathcal{J}}$.
		\end{itemize} 
	\end{lemma}

	\begin{proof}
		Following Observation \ref{fact:location}, two bursts of $(t_1,t_2)$-DI in $\mathbf{x}$ induce two bursts of $(1,t'-1)$-DS in $X[1]$. Let $X'_{t_1-t_2}$ denote the matrix representation of $\mathbf{x}'$. Hence, $\mathbf{x}'\in\mathcal{B}_{2,(t_1,t_2)}^{DI}(\mathbf{x})$ and $X'[1]\in\mathcal{B}_{2,(1,t'-1)}^{DS}(X[1])$.
		According to Lemma \ref{thm:Tt}, $X[1]$ can be recovered by $f(X[1])$ and $X'[1]$.
		Moreover,  since $X[1]$ is a $d$-regular sequence of length $n/(t_1-t_2)$, we distinguish two cases following Lemma  \ref{lem:2(1,t')-DS}.
		
	\textbf{Case i) Lemma  \ref{lem:2(1,t')-DS}-a) holds.} Denoting 
		\begin{align*}
			\mathcal{I}_{1}=[i,i'],\mathcal{I}_2=[j,j']\subseteq\left[{n\over t_1-t_2}\right]
		\end{align*}
		with 
		$$|\mathcal{I}_1|,|\mathcal{I}_2|\le d\log \left({n\over t_1-t_2}\right),$$
		by Observation \ref{fact:location},  two bursts of $(t_1,t_2)$-DI occur in $\mathbf{x}_{\mathcal{J}_1}$ and $\mathbf{x}_{\mathcal{J}_2}$ with
		\begin{align*}
			\mathcal{J}_1=\left[(i-t'-1)(t_1-t_2)+2,(i'+t'-1)(t_1-t_2)\right],\\
			\mathcal{J}_2=\left[(j-t'-1)(t_1-t_2)+2,(j'+t'-1)(t_1-t_2)\right].
		\end{align*}
		It is clear that 
		$$|\mathcal{J}_1|,|\mathcal{J}_2|\le \left(d\log \left({n\over t_1-t_2}\right)+2t'+1\right)(t_1-t_2).$$
		
	\textbf{ Case ii) Lemma  \ref{lem:2(1,t')-DS}-b) holds.} Denoting $$\mathcal{I}=[i,i']\subseteq\left[{n\over t_1-t_2}\right],|\mathcal{I}|\le 3d\log \left({n\over t_1-t_2}\right)+2t',$$
		 two bursts of $(t_1,t_2)$-DI occur in $\mathbf{x}_{\mathcal{J}}$ with
		$$\mathcal{J}=\left[(i-t'-1)(t_1-t_2)+2,(i'+t'-1)(t_1-t_2)\right].$$ 
		It is clear that 
		$$|\mathcal{J}|\le \left(3d\log \left({n\over t_1-t_2}\right)+4t'+1\right)(t_1-t_2).$$
		
		This completes the proof.
	\end{proof}
	
	After approximately locating the positions of two bursts of $(t_1,t_2)$-DI, in what follows, we present an approach to recover the original sequence with the help of Theorem \ref{thm:Tts1}.

\begin{figure}[t]
	\begin{center}
		\includegraphics[scale=0.25]{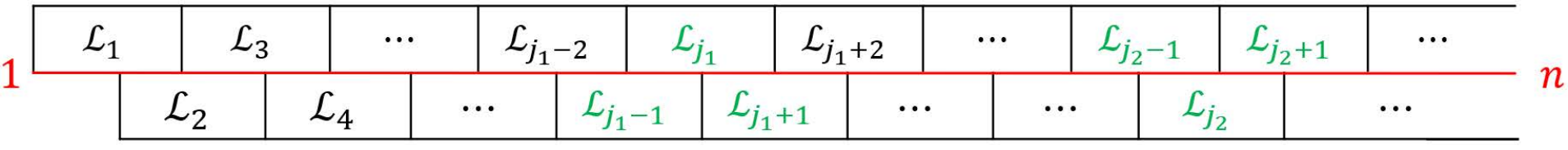}
		\caption{The  intervals $\mathcal{L}_j,j\in[\lceil n/\rho\rceil-1]$ in \eqref{eqn:interval}.}
		\label{fig:j1j2-1}
	\end{center}
\end{figure}

	Let $\rho$ be a positive integer, and $\mathcal{L}_j,j\in[\lceil n/\rho\rceil-1],$ be intervals of length $2\rho$, where
	\begin{align}\label{eqn:interval}
	\hspace{-6pt}	\mathcal{L}_j=\left\{\begin{array}{ll}
			[(j-1)\rho+1,(j+1)\rho], & 	\hspace{-8pt}\text{for~~} j\in[\lceil n/\rho\rceil -2],\\
			{[(j-1)\rho+1,n]},& 	\hspace{-8pt}\text{for~~} j=\lceil n/\rho\rceil -1.
		\end{array}\right.
	\end{align}
	The intervals defined in \eqref{eqn:interval} are illustrated in Fig. \ref{fig:j1j2-1}.

	Clearly,  for any length-$\rho$ interval $\mathcal{L}\subseteq[n]$, there exists an $\mathcal{L}_{j_1},j_1\in[\lceil n/\rho\rceil -1],$ such that $\mathcal{L}\subseteq \mathcal{L}_{j_1}$.
%
	
	\begin{observation}\label{obser:intervals}
		As shown in Fig. \ref{fig:j1j2-1}, if there is one burst of $(t_1,t_2)$-DI in an interval $\mathcal{L}_j,j\in[2,\lceil n/\rho\rceil-2]$, the adjacent  intervals $\mathcal{L}_{j-1}$ and $\mathcal{L}_{j+1}$ might also be affected. In particular, for $j=1$ and $j=\lceil n/\rho\rceil-1$, only $\mathcal{L}_2$ and $\mathcal{L}_{\lceil n/\rho\rceil-2}$ may be affected, respectively.
	\end{observation}
	
	As a result, we provide a construction to correct two bursts of $(t_1,t_2)$-DI for binary sequences.

	\begin{construction}\label{con1}
		For any $\mathbf{x}\in\Sigma_2^n$ and two positive integers $t_1, t_2$ with $t_1>t_2$, denote by $X_{t_1-t_2}$ the matrix representation of $\mathbf{x}$ following \eqref{eqn:matrix}, and $X[1]$ the first row of $X_{t_1-t_2}$.
		Let $f_1(\mathbf{x}),f_2(\mathbf{x})$ and $f(\mathbf{x})$ be the functions capable of correcting one burst of $(t_1,t_2)$-DI, two bursts of $(t_1,t_2)$-DI, and two bursts of $(1,t'-1)$-DS proposed in Corollary \ref{coro}, Construction \ref{con2},   and Lemma \ref{thm:Tt} for any $\mathbf{x}\in\Sigma_2^n$, respectively.
		Let $\mathcal{L}_j$ and $\mathcal{L}_{j'}$ be the intervals defined in \eqref{eqn:interval} with $\rho_1=(d\log (n/(t_1-t_2))+2t'+1)(t_1-t_2)$ and $\rho_2=(3d\log (n/(t_1-t_2))+4t'+1)(t_1-t_2)$, respectively, where $j\in[\lceil n/\rho_1\rceil-1],j'\in[\lceil n/\rho_2\rceil-1]$. For each $\mathbf{x}\in\Sigma_2^n$ with $X[1]$ being a $d$-regular sequence, $a\in[0,2]$ and $i\in[0,1]$, denote
		\begin{align}\label{eqn:bar-phi}
			\phi^{(a)}_i(\mathbf{x})=\sum_{j\in[\lceil n/\rho_1\rceil-1],\atop j\equiv a \bmod 3} j^i\cdot f_1(\mathbf{x}_{\mathcal{L}_j}) \pmod{2n^iN_1}, 
		\end{align}
		\begin{align}\label{eqn:bar-h}
			h^{(i)}(\mathbf{x})=\sum_{j\in[\lceil n/\rho_1\rceil-1],\atop j\equiv i \bmod 2} f_2(\mathbf{x}_{\mathcal{L}_j}) \pmod{N_2}, 
		\end{align}
		and
		\begin{align}\label{eqn:bar-h1}
			g^{(i)}(\mathbf{x})=\sum_{j'\in[\lceil n/\rho_2\rceil-1],\atop j'\equiv i \bmod 2} f_2(\mathbf{x}_{\mathcal{L}_{j'}}) \pmod{N_3},
		\end{align}
		where 
		\begin{align*}
			N_1=2^{4\log (2\rho_1)+o(\log \rho_1)},\\
			N_2=2^{8\log (2\rho_1)+o(\log \rho_1)},
		\end{align*} 
and 
	\begin{align*}
		N_3=2^{8\log (2\rho_2)+o(\log \rho_2)}.
	\end{align*}
		
		Then, define
		\begin{align*}
			\psi(\mathbf{x})=\Big(f(X[1]), 	\phi^{(a)}_i(\mathbf{x}),h^{(i)}(\mathbf{x}),g^{(i)}(\mathbf{x}),\notag\\
			i\in[0,1],a\in[0,2]\Big).
		\end{align*}

	For some fixed parameters 
	\begin{align*}
		f\in\big[0,2^{8\log (n/(t_1-t_2))+o(\log n)}\big),
	\end{align*}
$\phi_{i,a}\in[0,2n^iN_1)$, $h_i\in[0,N_2)$,
	 and $g_i\in[0,N_3)$, $a\in[0,2]$, $i\in[0,1]$, respectively,
	define a code by
	\begin{align*}
		\mathcal{C}_2\triangleq\Big\{&\mathbf{x}\in\Sigma_2^n:\text{ $X[1]$ is a d-regular sequence},\\
		&\psi(\mathbf{x})=(f,\phi_{i,a},h_i,g_i),i\in[0,1],
		a\in[0,2]\Big\}.
	\end{align*}
	\end{construction}

	\begin{theorem}\label{thm:t1>t2}
	The code $\mathcal{C}_2$ generated in	Construction  \ref{con1} can correct two bursts of $(t_1,t_2)$-DI  with at most $11\log n+o(\log n)$ bits of redundancy. 
	\end{theorem}
	\begin{proof}
		Based on the function $f(X[1])$ and $X'[1]\in\mathcal{B}_{2,(1,t'-1)}^{DS}(X[1])$, $X[1]$ can be recovered and the approximate locations of two bursts of $(t_1,t_2)$-DI can be identified following Lemmas \ref{thm:Tt} and \ref{thm:2bursts}. Then,	we prove this result by considering the following two cases.

		\textbf{Case 1.} Lemma \ref{thm:2bursts}-a) holds. In this case, we can find two intervals $\mathcal{J}_1$ and $\mathcal{J}_2$ with $|\mathcal{J}_1|,|\mathcal{J}_2|\le \rho_1= (d\log (n/(t_1-t_2))+2t'+1)(t_1-t_2)$ such that two bursts occur in $\mathbf{x}_{\mathcal{J}_1}$ and $\mathbf{x}_{\mathcal{J}_2}$, respectively.
		Following \eqref{eqn:interval}, there exist two intervals $\mathcal{L}_{j_1},\mathcal{L}_{j_2}$ such that $\mathcal{J}_1\subseteq \mathcal{L}_{j_1},\mathcal{J}_2\subseteq \mathcal{L}_{j_2}$, where $j_1,j_2\in[\lceil n/\rho_1\rceil-1]$.
		Without loss of generality, let $j_1\le j_2$ in the following.
		
		We consider the following two subcases:
		
		\textit{1) $j_1,j_2\in[2,\lceil n/\rho_1\rceil-2]$ with $j_1\ne j_2$.}	
			Assume that $\mathcal{L}_{j_1}=[\lambda_0,\lambda_1]$, $\mathcal{L}_{j_2}=[\lambda_2,\lambda_3]$.
		Thus, we have
		\begin{align}\label{eqn:lambda}
			&\mathbf{x}_{[1,\lambda_0)}=\mathbf{x}'_{[1,\lambda_0)}, \notag\\ &\mathbf{x}_{[\lambda_1+1,\lambda_2)}=\mathbf{x}'_{[\lambda_1-t_1+t_2+1,\lambda_2-t_1+t_2)},\\
			&\mathbf{x}_{[\lambda_3+1,n]}=\mathbf{x}'_{[\lambda_3-2t_1+2t_2+1,n-2t_1+2t_2]},\notag
		\end{align}
		and 
		\begin{align}\label{eqn:x'x}
			\begin{array}{lll}
			\mathbf{x}'_{[\lambda_0,\lambda_1-t_1+t_2]}\in\mathcal{B}_{(t_1,t_2)}^{DI}(\mathbf{x}_{\mathcal{L}_{j_1}}),\\
			\mathbf{x}'_{[\lambda_2-t_1+t_2,\lambda_3-2t_1+2t_2]}\in\mathcal{B}_{(t_1,t_2)}^{DI}(\mathbf{x}_{\mathcal{L}_{j_2}}).
			\end{array}
		\end{align}
		
		Denoting $j_1 \bmod 3 =m_1$ and $j_2\bmod 3 = m_2$,	
		according to Observation \ref{obser:intervals} and \eqref{eqn:bar-phi}, for any $a\in[0,2],i\in[0,1]$, we can obtain 
		
		\textit{i)} when $a=m_1=m_2$, 
		\begin{align}\label{m1m2l}
			&j_1^i\cdot f_1(\mathbf{x}_{\mathcal{L}_{j_1}})+j_2^i\cdot f_1(\mathbf{x}_{\mathcal{L}_{j_2}})\notag\\
			=&
			\phi^{(a)}_i(\mathbf{x})-\sum_{j\equiv a \bmod 3,\atop j\notin\{j_1,j_2\}} j^i\cdot f_1(\mathbf{x}_{\mathcal{L}_j}) \pmod{2n^iN_1}.
		\end{align}
		Note that since it is evident that  $j\in[\lceil n/\rho_1\rceil-1]$,   we omit this condition hereafter for notational simplicity.

		\textit{ii)} when $a=m_1\equiv m_2-1\pmod 3$, 
		\begin{align}\label{m1m2-1l}
			&j_1^i\cdot f_1(\mathbf{x}_{\mathcal{L}_{j_1}})+(j_2-1)^i\cdot f_1(\mathbf{x}_{\mathcal{L}_{j_2-1}})\notag\\
			=&
			\phi^{(a)}_i(\mathbf{x})-\sum_{j\equiv a \bmod 3,\atop j\notin\{j_1,j_2-1\}} j^i\cdot f_1(\mathbf{x}_{\mathcal{L}_j}) \pmod{2n^iN_1},
		\end{align}
		and
		\begin{align}\label{m1+1m2l}
			&(j_1+1)^i\cdot f_1(\mathbf{x}_{\mathcal{L}_{j_1+1}})+j_2^i\cdot f_1(\mathbf{x}_{\mathcal{L}_{j_2}})\notag\\
			=&
			\phi^{(a+1)}_i(\mathbf{x})-\sum_{j\equiv a \bmod 3,\atop j\notin\{j_1+1,j_2\} } j^i\cdot f_1(\mathbf{x}_{\mathcal{L}_j}) \pmod{2n^iN_1}.
		\end{align}
		Note that the superscript $a +1$ of $\phi_i$ is taken modulo $3$.

		\textit{iii)} when $a=m_1\equiv m_2-2\pmod 3$,
		\begin{align}\label{m1m2-2l}
			&j_1^i\cdot f_1(\mathbf{x}_{\mathcal{L}_{j_1}})+(j_2+1)^i\cdot f_1(\mathbf{x}_{\mathcal{L}_{j_2+1}})\notag\\=
			&
			\phi^{(a)}_i(\mathbf{x})-\sum_{j\equiv a \bmod 3,\atop j\notin\{j_1,j_2+1\} } j^i\cdot f_1(\mathbf{x}_{\mathcal{L}_j}) \pmod{2n^iN_1}, 
		\end{align}
		and
		\begin{align}\label{m1-1m2l}
		&	(j_1-1)^i\cdot f_1(\mathbf{x}_{\mathcal{L}_{j_1-1}})+j_2^i\cdot f_1(\mathbf{x}_{\mathcal{L}_{j_2}})\notag\\
		=&
			\phi^{(a+2)}_i(\mathbf{x})-\sum_{j\equiv a \bmod 3,\atop j\notin\{j_1-1,j_2\}} j^i\cdot f_1(\mathbf{x}_{\mathcal{L}_j}) \pmod{2n^iN_1}.
		\end{align}
		Note that the superscript $a+2$ of $\phi_i$ is taken modulo $3$.
		
		By Corollary \ref{coro}, we know 
		$f_1(\mathbf{x}_{\mathcal{L}_j})\le N_1,j\in[\lceil n/\rho_1\rceil-1]$. Therefore, regardless of which of the three subcases above applies,  $f_1(\mathbf{x}_{\mathcal{L}_{j_1}})$ and $f_1(\mathbf{x}_{\mathcal{L}_{j_2}})$ can be determined from \eqref{m1m2l}-\eqref{m1-1m2l}.  
		Since two bursts of $(t_1,t_2)$-DI are in $\mathbf{x}_{\mathcal{L}_{j_1}}$ and $\mathbf{x}_{\mathcal{L}_{j_2}}$, 
		the original sequence 
		$\mathbf{x}$ can be uniquely recovered following \eqref{eqn:lambda}, \eqref{eqn:x'x}, and Theorem \ref{thm:Tts1}.
		
		In particular,	for the case of $j_1=1$, Equation \eqref{m1-1m2l} shows
		\begin{align*}
		\scalebox{0.85}{$
			\begin{aligned}
		j_2^i\cdot f_1(\mathbf{x}_{\mathcal{L}_{j_2}})=
			\phi^{(a+2)}_i(\mathbf{x})-\sum_{j\equiv a \bmod 3,
				\atop j\ne j_2} j^i\cdot f_1(\mathbf{x}_{\mathcal{L}_j}) \pmod{2n^iN_1}
		\end{aligned}$}
	\end{align*}
		and for the case of $j_2=\lceil n/\rho_1\rceil-1$, Equation \eqref{m1m2-2l} gives
		\begin{align*}
			\scalebox{0.85}{$
				\begin{aligned}
			j_1^i\cdot f_1(\mathbf{x}_{\mathcal{L}_{j_1}})=
			\phi^{(a)}_i(\mathbf{x})-\sum_{j\equiv a \bmod 3,\atop j\ne j_1 } j^i\cdot f_1(\mathbf{x}_{\mathcal{L}_j}) \pmod{2n^iN_1}.
		\end{aligned}$}
	\end{align*}
		This means that there are also at most two unknowns in equations \eqref{m1m2l}-\eqref{m1-1m2l}, which can be solved accordingly.
		
		\textit{2) $j_1,j_2\in[\lceil n/\rho_1\rceil-1]$ with $j_1= j_2$.}	
		In this subcase, two bursts of $(t_1,t_2)$-DI occur in a length-$2\rho_1$ substring $\mathbf{x}_{\mathcal{L}_{j_1}}$ of $\mathbf{x}$. Assume that $\mathcal{L}_{j_1}=[\lambda_0,\lambda_1]$, thus,
		\begin{align}\label{eqn:lambda2}
			\begin{array}{lll}
			\mathbf{x}_{[1,\lambda_0)}=\mathbf{x}'_{[1,\lambda_0)}, \\
			\mathbf{x}_{[\lambda_1+1,n]}=\mathbf{x}'_{[\lambda_1-2t_1+2t_2+1,n-2t_1+2t_2]},
		\end{array}\end{align}
		and 
		\begin{align}\label{eqn:x'x2}
			\mathbf{x}'_{[\lambda_0,\lambda_1-2t_1+2t_2]}\in\mathcal{B}_{2,(t_1,t_2)}^{DI}(\mathbf{x}_{\mathcal{L}_{j_1}}).
		\end{align}
		
		According to \eqref{eqn:bar-h}, 	when $j_1\bmod 2=0$, we can compute
		\begin{align}\label{eqn:t1t2}
			f_2(\mathbf{x}_{\mathcal{L}_{j_1}})=h^{(0)}(\mathbf{x})-\sum_{j\bmod 2=0,\atop j\ne j_1} f_2(\mathbf{x}_{\mathcal{L}_j}) \pmod{N_2},
		\end{align}
		whereas, when $j_1\bmod 2=1$,
		\begin{align}\label{eqn:t1t21}
			f_2(\mathbf{x}_{\mathcal{L}_{j_1}})=h^{(1)}(\mathbf{x})-\sum_{j\bmod 2=1,\atop j\ne j_1} f_2(\mathbf{x}_{\mathcal{L}_j}) \pmod{N_2}.
		\end{align}
		
		From Theorem \ref{thm:Tts1}, we know 
		$f_2(\mathbf{x}_{\mathcal{L}_j})\le N_2,j\in[\lceil n/\rho_1\rceil-1]$.
		Since  $\mathbf{x'}\in \mathcal{B}_{2,(t_1,t_2)}^{DI}(\mathbf{x})$,
		using $f_2(\mathbf{x}_{\mathcal{L}_{j_1}})$ in \eqref{eqn:t1t2} or \eqref{eqn:t1t21},
		the original sequence
		$\mathbf{x}$ can be uniquely recovered following \eqref{eqn:lambda2}, \eqref{eqn:x'x2},  and Theorem \ref{thm:Tts1}.
		
		\textbf{Case 2.} Lemma \ref{thm:2bursts}-b) holds. In this case, we can find an interval $\mathcal{J}\subseteq[n]$ with $|\mathcal{J}|\le \rho_2=(3d\log (n/(t_1-t_2))+4t'+1)(t_1-t_2)$ such that 
	 two bursts of $(t_1,t_2)$-DI occur in $\mathbf{x}_{\mathcal{J}}$.
		By \eqref{eqn:interval}, there exists an interval $\mathcal{L}_{j}$ such that $\mathcal{J}\subseteq \mathcal{L}_{j}$, where $j\in[\lceil n/\rho_2\rceil-1]$.
		Then, the proof is similar to that of subcase 2) in Case 1, with the only difference being the use of $g^{(i)}(\mathbf{x})$ in place of $h^{(i)}(\mathbf{x})$. The details are therefore omitted.

		Using Lemma \ref{thm:Tt} and \eqref{eqn:bar-phi}-\eqref{eqn:bar-h1}, the length of $\psi(\mathbf{x})$ satisfies
		\begin{align*}
			\scalebox{0.9}{$\begin{aligned}
			\left|\psi(\mathbf{x})\right|=&\left|f(X[1])\right|+\sum_{i=0}^1\sum_{a=0}^2\left|\phi^{(a)}_i(\mathbf{x})\right|+\sum_{i=0}^1\left|h^{(i)}(\mathbf{x})\right|\\
			&+\sum_{i=0}^1\left|g^{(i)}(\mathbf{x})\right|\\
			=&8\log \left({n\over t_1-t_2}\right)+o(\log n) +
			3\left(\log(2N_1)+\log(2nN_1)\right)\\
			&+2(\log N_2)+2(\log N_3)\\
			=&11\log{n}+o(\log n).
		\end{aligned}$}\end{align*}
	
From Lemma \ref{lem:regular}, for $\mathbf{x}\in\Sigma_2^n$ with $X_{t_1-t_2}$ being its matrix representation, encoding $X[1]$ into a $d$-regular sequence requires only one additional redundancy bit. Therefore, the $d$-regular constraint introduces only a constant redundancy overhead and does not affect the asymptotic order of resulting redundancy.
	Consequently, the total redundancy  of $\mathcal{C}_2$ remains $11\log{n}+o(\log n)$.
	This completes the proof.
	\end{proof}

	
	According to Lemma \ref{lem:regular}  and Theorems \ref{thm:t1=t2}, \ref{thm:t1>t2}, the following conclusion is straightforward.
	\begin{theorem}\label{thm:t1t2}
		For any fixed positive integers $t_1,t_2$ with $t_1\ge t_2$, there exists a family of  binary codes correcting two bursts of $(t_1,t_2)$-DI errors with redundancy at most $11\log n+o(\log n)$.
	\end{theorem}

		\subsection{Complexity}
We compare the computational complexity of the proposed schemes in this section with the one obtained by the syndrome compression technique directly in Section \ref{sec:syndrome}.
	As shown in \cite{sima2020syndrome},
	the main computational cost of the syndrome compression technique arises from a brute-force search for a parameter
	\begin{align*}
		a(\mathbf{x})\le 2^{\log{\left|\mathcal{N}^{*}\right|+o(\log n)}}
	\end{align*} such that 
\begin{align*}
	a(\mathbf{x}) \notin\big\{j:j\mid \big(h(\mathbf{x})-h(\mathbf{x}')\big)\big\}
\end{align*} for any $\mathbf{x}'\in\mathcal{N}^{*}$, where 
\begin{align*}
\mathcal{N}^{*}\in\left\{\mathcal{N}^{DI}_{2,(t_1,t_2)}(\mathbf{x}),\mathcal{N}^{DS}_{2,(1,t'-1)}(X[1])\right\}
\end{align*} is the error type in this paper and $h(\mathbf{x})$ is defined in Lemma \ref{lemma:DS}.
	Therefore, for syndrome-compression-based constructions, the computational complexity is mainly determined by the size of 
	$\mathcal{N}^{*}$, whereas for constructions based on RS codes, it is determined by the complexity of the corresponding encoding and decoding algorithms.
	
						\begin{lemma}\label{lem:complexity}
			The computational complexity of the proposed schemes in Theorem \ref{thm:t1t2} is $O(n\log n)$ when $t_1=t_2$ and $O\big(\big(n/(t_1-t_2)\big)^4\big)$ when $t_1>t_2$, while that of the codes in Theorem \ref{thm:Tts1} is $O(n^4)$.
		\end{lemma}	
		\begin{proof}
			For the codes in Theorem \ref{thm:t1t2}, 
			when $t_1=t_2$, two bursts of $(t_1,t_2)$-DI errors are converted to four substitutions, which are corrected by a coset of an RS code as in Theorem \ref{thm:t1=t2}. Both the  encoding and decoding complexities for RS codes are $O(n\log n)$.
			
			When $t_1>t_2$, two bursts of $(1,t'-1)$-DS in $X[1]$ are first identified using a function obtained by the syndrome compression technique, and then the original error pattern in $\mathbf{x}$, i.e., two bursts of $(t_1,t_2)$-DI, are corrected.
			Since the length of $X[1]$  is $ n/ (t_1-t_2)$, following \eqref{eqn:Nt1}, we can compute
			\begin{align}\label{eqn:con1}
					\scalebox{0.82}{$
					\begin{aligned}
						&\left|\mathcal{N}^{DS}_{2,(1,t'-1)}(X[1])\right|\\
						\le&4\cdot {\lceil{n\over t_1-t_2}\rceil-2t'+2\choose 2}{\lceil{n\over t_1-t_2}\rceil-2t'+3\choose 2}\Big(\sum_{i=0}^{t'-2}{t'-2\choose i}\Big)^4\\
						=& O\left(\left({n\over {t_1-t_2}}\right)^4\right).
					\end{aligned}	
						$}
			\end{align}
			
			In addition, the functions $\phi^{(a)}_i(\mathbf{x})$, $h^{(i)}(\mathbf{x})$ and $g^{(i)}(\mathbf{x})$ in Construction \ref{con1} can be computed in $O(n\rho_1)$, $O(n\rho_1^3)$ and $O(n\rho_2^{3})$ time, respectively. This follows from the facts that the computational complexities of $f_1(\mathbf{x}_{\mathcal{L}_j})$, $f_2(\mathbf{x}_{\mathcal{L}_j})$ and $f_2(\mathbf{x}_{\mathcal{L}_{j'}})$ are $O(\rho_1^2)$, $O(\rho_1^4)$ and $O(\rho_2^4)$, respectively, and that there are approximately $n/\rho_1$ summation terms in \eqref{eqn:bar-phi} and \eqref{eqn:bar-h}, and $n/\rho_2$ terms in \eqref{eqn:bar-h1}, where 
			\begin{align*}
				\rho_1=(d\log (n/(t_1-t_2))+2t'+1)(t_1-t_2),
			\end{align*} 
			and 
			\begin{align*}
				\rho_2=(3d\log (n/(t_1-t_2))+4t'+1)(t_1-t_2).
			\end{align*}
			This means that the functions $\phi^{(a)}_i(\mathbf{x}),h^{(i)}(\mathbf{x})$ and $g^{(i)}(\mathbf{x})$ have $O(n\log^3 n)$ time complexity.
			Hence, the computational complexity of Construction \ref{con1} is dominated by the computation of $f(X[1])$, i.e., the cardinality of $\mathcal{N}^{DS}_{2,(1,t'-1)}(X[1])$, which is $O((n/(t_1-t_2))^4)$ from \eqref{eqn:con1}.
			
			Regarding the codes developed in Theorem \ref{thm:Tts1}, the computational complexity arises from $|\mathcal{N}_{2,(t_1,t_2)}^{DI}(\mathbf{x})|$ since they are obtained by using the syndrome compression technique directly, which is $O(n^4)$ as shown in Lemma \ref{lem:NDI}.
		\end{proof}

		\begin{table}[t]
		\centering
		\begin{center}\caption{Comparison of the codes in Thm. \ref{thm:Tts1} and Thm. \ref{thm:t1t2} under some fixed parameters.}\label{comparison}
			\renewcommand\arraystretch{1.5}	
			\resizebox{0.485\textwidth}{!}{\begin{tabular}{|c|c|c|c|c|c|c|c|}
				\hline
				\multirow{2}{*}{Row}  &	\multirow{2}{*}{Code length $n$}  & \multirow{2}{*}{$(t_1,t_2)$-DI} & \multirow{2}{*}{$t'$}  & \multicolumn{2}{c|}{$\left|\mathcal{N}^{*}\right|$}  \\
				\cline{5-8}
				&&&& $|\mathcal{N}_{2,(t_1,t_2)}^{DI}(\mathbf{x})|$&$|\mathcal{N}_{2,(1,t'-1)}^{DS}(X[1])|$\\
				\hline
				1&$1024$ & \multirow{3}{*}{$(10,8)$} &\multirow{3}{*}{$5$} &  $2^{66}$ & $2^{48}$\\
				\cline{1-2}\cline{5-6}
				2&$512$&  &  &   $2^{62}$ & $2^{44}$ \\
				\cline{1-2}\cline{5-6}
				3&$256$ &   & & $2^{58}$ & $2^{40}$ \\
				\cline{1-4}\cline{5-6}
				4&$1024$ & \multirow{3}{*}{$(10,7)$} &\multirow{3}{*}{$4$} &   $2^{64}$ & $2^{42}$\\
				\cline{1-2}\cline{5-6}
				5&$512$&  &  &   $2^{60}$ & $2^{38}$ \\
				\cline{1-2}\cline{5-6}
				6&$256$ &   & &  $2^{56}$ & $2^{34}$ \\
				\cline{1-4}\cline{5-6}
				7&$1024$ & \multirow{3}{*}{$(10,6)$}	   &\multirow{3}{*}{$3$}  &   $2^{62}$& $2^{36}$ \\
				\cline{1-2}\cline{5-6}
				8&$512$&    & &  $2^{58}$& $2^{32}$\\
				\cline{1-2}\cline{5-6}
				9&$256$& 	   & &  $2^{54}$ & $2^{28}$\\	
				\cline{1-4}\cline{5-6}
				10&$1024$& \multirow{3}{*}{$(10,1)$}	   &\multirow{3}{*}{$2$}&     $2^{54}$& $2^{28}$\\	
				\cline{1-2}\cline{5-6}
				11&$512$&    & &   $2^{50}$&$2^{24}$ \\
				\cline{1-2}\cline{5-6}
				12&$256$& 	   & &  $2^{46}$& $2^{20}$\\
				\hline
				13&\multicolumn{3}{|c|}{} & &Operations of RS codes \\ 
				\cline{1-4}\cline{5-6}
				14&	$512$ &$(10,10)$ & \multirow{3}{*}{-}&   $2^{74}$ & $2^{15}$\\
				\cline{1-3}\cline{5-6}
				15&	$512$& $(5,5)$   & &   $2^{54}$& $2^{17}$  \\
				\cline{1-3}\cline{5-6}
				16&	$512$& 	$(2,2)$   & & $2^{34}$& $2^{19}$  \\
				\hline
			\end{tabular}}
		\end{center}
	\end{table}
	
		\begin{remark}\label{remark}
			For fixed $t_1$ and $t_2$, when $t_1>t_2$, the proposed construction reduces the search space and therefore the constant factor of the brute-force search, while the asymptotic complexity remains $O(n^4)$. When $t_1=t_2$, the computational complexity is reduced to $O(n\log n)$ due to the application of RS codes.
		\end{remark}
		
	It is evident from Lemma \ref{lem:complexity} and Remark \ref{remark} that the codes in Construction \ref{con1} reduce the computational complexity by a factor of $(t_1-t_2)^4$.
	For clarity, we compare the upper bounds on $\mathcal{N}^*$ for Constructions \ref{con2} and \ref{con1} by a specific example. Note that the numerical values below are not exact, since they are evaluated for finite $n$, while the complexity analysis is asymptotic in nature (i.e., assumes $n\rightarrow \infty$).  Nevertheless, we provide them to elucidate the computational complexity benefits of the new constructions. 
	
	For $n=256, t_1=10,t_2=2$, we have $t'=2$. For any binary sequence $\mathbf{x}\in\Sigma_2^n$, if two bursts of $(t_1,t_2)$-DI occur in $\mathbf{x}$, from \eqref{eqn:con2} we have 
\begin{align}\label{N-con2}
		|\mathcal{N}_{2,(10,2)}^{DI}(\mathbf{x})|
		\le{238\choose 2}{241\choose 2}\times 2^{16}\approx 2^{46},
	\end{align}
whereas \eqref{eqn:con1} gives
	\begin{align}\label{N-con1}
		|\mathcal{N}^{DS}_{2,(1,1)}(X[1])|\le4\cdot {30\choose 2}{31\choose 2}\approx 2^{22}.
	\end{align}
	
	It is clear that the computational complexity associated with \eqref{N-con1} is significantly lower than that of \eqref{N-con2}, indicating that
	 the  practical applicability of Construction \ref{con2} is limited due to the substantially higher computational cost. In contrast, Construction \ref{con1} exhibits superior search efficiency and is therefore more amenable to implementation.
%

	Furthermore, a numerical comparison between the codes in Theorems \ref{thm:Tts1} and \ref{thm:t1t2} for several fixed parameter settings is presented in Table \ref{comparison}.
	It can be observed that, across all the considered parameter settings, the brute-force search complexity of the construction in Theorem \ref{thm:t1t2} is much lower than that of the construction in Theorem \ref{thm:Tts1}. For instance, when $n=1024$, $512$, and $256$, if two bursts of $(t_1,t_2)=(10,6)$-DI errors occur in a binary sequence $\mathbf{x}\in\Sigma_2^n$, the construction in Theorem \ref{thm:t1t2} replaces the brute-force search over $\mathcal{N}^{DI}_{2,(10,6)}(\mathbf{x})$ with a search over the substantially smaller set $\mathcal{N}^{DS}_{2,(1,2)}(X[1])$. Specifically, the upper bound on the search space is reduced from $\left|\mathcal{N}^{DI}_{2,(10,6)}(\mathbf{x})\right|\approx 2^{62},\,2^{58},\,\text{and }2^{54}$ in Theorem \ref{thm:Tts1} to $\left|\mathcal{N}^{DS}_{2,(1,2)}(X[1])\right|\approx 2^{36},\,2^{32},\,\text{and }2^{28}$, respectively, corresponding to a reduction by a factor of $2^{26}$ for all three code lengths (Rows 7-9).


\begin{figure}[ht]
		\begin{center}
			\includegraphics[scale=0.5]{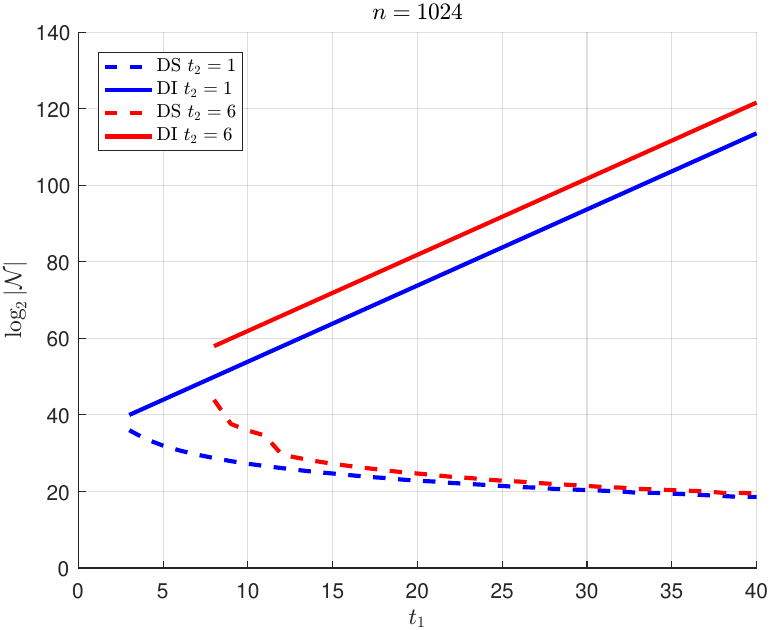}
			\caption{Comparison of $\mathcal{N}^{DI}(\mathbf{x})$ and  $\mathcal{N}^{DS}(X[1])$  for varying $t_1$ with $t_2=1,6$ and $n=1024$.}
			\label{fig:t1}
		\end{center}
	\end{figure}
	\begin{figure}[ht]
		\begin{center}
			\includegraphics[scale=0.45]{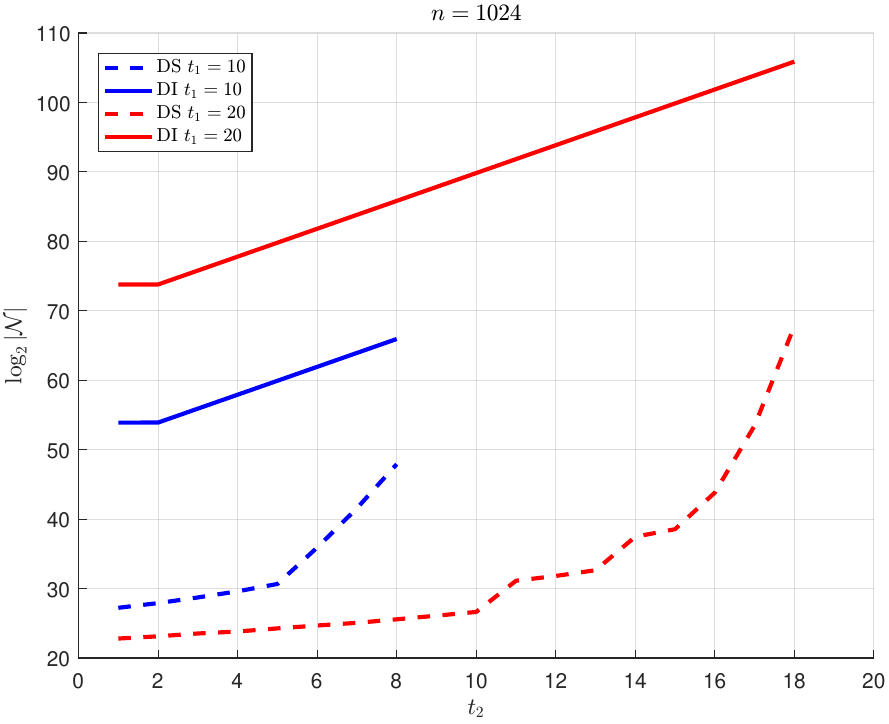}
			\caption{Comparison of $\mathcal{N}^{DI}(\mathbf{x})$ and  $\mathcal{N}^{DS}(X[1])$  for varying $t_2$  with $t_1=10,20$  and $n=1024$.}
			\label{fig:t2}
		\end{center}
	\end{figure}\begin{figure}[ht]
		\begin{center}
			\includegraphics[scale=0.6]{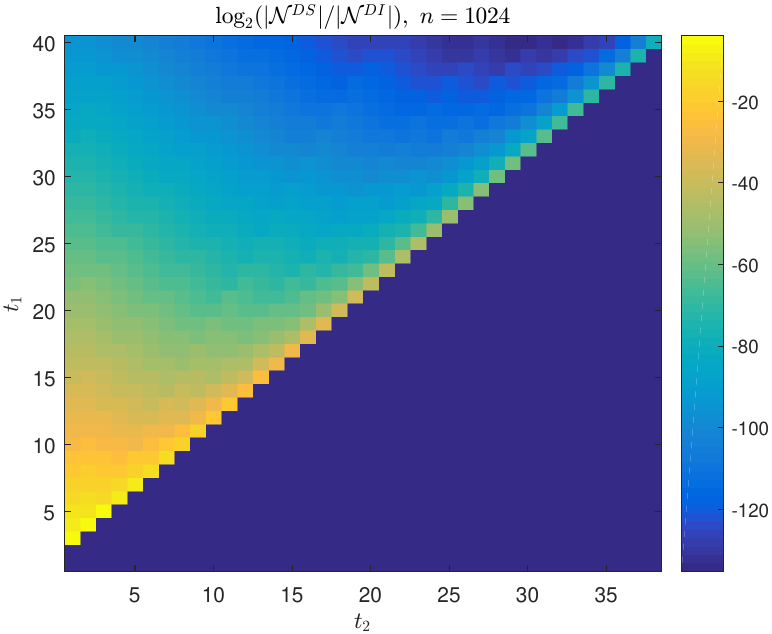}
			\caption{Comparison of $\mathcal{N}^{DI}(\mathbf{x})$ and  $\mathcal{N}^{DS}(X[1])$  for varying $t_1$ and $t_2$   with $n=1024$.}
			\label{fig:t1t2}
		\end{center}
	\end{figure}\begin{figure}[ht]
		\begin{center}
			\includegraphics[scale=0.55]{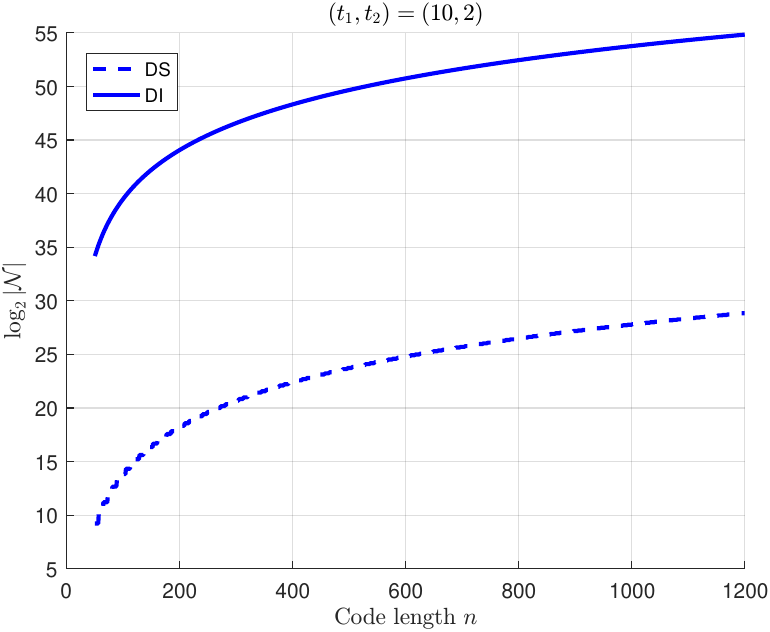}
			\caption{Comparison of $\mathcal{N}^{DI}(\mathbf{x})$ and  $\mathcal{N}^{DS}(X[1])$  for varying $n$ with $t_1=10$  and $t_2=2$.}
			\label{fig:n}
		\end{center}
	\end{figure}

	To illustrate the computational complexity more explicitly,
	we compare the upper bounds of $|\mathcal{N}^{DI}_{2,(t_1,t_2)}(\mathbf{x})|$ and $|\mathcal{N}^{DS}_{2,(1,t'-1)}(X[1])|$, which are shown in Equations \eqref{eqn:con2}, \eqref{eqn:con21}, and \eqref{eqn:con1}, respectively, in Fig. \ref{fig:t1}–\ref{fig:t1t2} for a fixed code length $n=1024$, and in Fig. \ref{fig:n} for fixed parameters $t_1=10$ and $t_2=2$. 
	Since $|\mathcal{N}^*|$ grows exponentially with respect to $t_1$ and $t_2$, all results are plotted in the log domain.
	Note that the subscripts of $\mathcal{N}^*$ are omitted whenever the context is clear.
	
	Fig. \ref{fig:t1} illustrates the dependence of $|\mathcal{N}^*|$ on $t_1$ for fixed $t_2$.
	It can be observed that $|\mathcal{N}^{DI}(\mathbf{x})|$ increases rapidly with $t_1$, exhibiting an approximately linear growth in the log scale. In contrast, $|\mathcal{N}^{DS}(X[1])|$ decreases and eventually stabilizes. Consequently, the gap between the two quantities widens as $t_1$ increases. This indicates that two-burst $(t_1,t_2)$-DI ECCs obtained via the syndrome compression technique directly have substantially higher brute-force search complexity than the resulting codes in Construction \ref{con1}, particularly when the burst-length $t_1$ is large.  
	
	Fig. \ref{fig:t2} shows the effect of $t_2$ on $|\mathcal{N}^*|$ for fixed $t_1$.
	When $t_2$ increases, both set sizes grow; however, $|\mathcal{N}^{DI}(\mathbf{x})|$ increases at a substantially faster rate. The DS curves remain relatively flat over a wide range of $t_2$ and consistently stay below the DI curves.
	This behavior confirms that  two-burst $(t_1,t_2)$-DI ECCs obtained by directly applying the syndrome compression technique inherently induce higher complexity.
	
	Fig. \ref{fig:t1t2} provides a global comparison over all considered pairs $(t_1,t_2)$.
	The value of $\log_2(|\mathcal{N}^{DS}(X[1])|/|\mathcal{N}^{DI}(\mathbf{x})|)$ remains negative for the entire parameter range, demonstrating that
$|\mathcal{N}^{DI}(\mathbf{x})| > |\mathcal{N}^{DS}(X[1])|$. Furthermore, the magnitude of the difference grows as $t_1$ increases or $t_2$ decreases, revealing a uniform search-space advantage of the DS-based formulation across the parameter space.

	Fig. \ref{fig:n} studies the scaling behavior of $|\mathcal{N}^*|$ with respect to the code length $n$.
	When $t_1=10$ and $t_2=2$ are fixed, both $|\mathcal{N}^{DS}(X[1])|$ and $|\mathcal{N}^{DI}(\mathbf{x})|$ increase with $n$, but the curve corresponding to $|\mathcal{N}^{DI}(\mathbf{x})|$ consistently remains above that of $|\mathcal{N}^{DS}(X[1])|$, suggesting a higher complexity.

	Overall, the four figures consistently demonstrate that two-burst $(t_1,t_2)$-DI ECCs obtained by directly applying  the syndrome compression technique incur significantly higher brute-force search complexity than the resulting codes in Construction \ref{con1} across the entire parameter ranges and code lengths considered. 
	This observation provides strong evidence for the lower complexity of the proposed constructions.

	\section{Conclusions}\label{sec:conclusions}
	In this paper, we developed binary ECCs capable of correcting two bursts of $(t_1,t_2)$-DI errors. We first established that two-burst $(t_1,t_2)$-DI ECCs, two-burst $(t_2,t_1)$-DI ECCs, and one-burst $(t_1,t_2)$-DI together with one-burst $(t_2,t_1)$-DI ECCs are equivalent. Leveraging this equivalence, we characterized lower and upper bounds on the code size of two-burst  $(t_1,t_2)$-DI ECCs, thereby showing that the minimum redundancy of such codes lies between  $2\log n+O(1)$ and $4\log n +o(\log n)$.
	Furthermore, focusing on the cases with $t_1=t_2$ and $t_1 > t_2$, we presented constructions of binary ECCs correcting two bursts of $(t_1,t_2)$-DI errors with redundancy at most $8\log n+o(\log n)$ and $11\log n + o(\log n)$, respectively. The latter was obtained by transforming two bursts of $(t_1,t_2)$-DI errors in a binary sequence into two bursts of $(1,t'-1)$-DS errors in the first row of its matrix representation, coupled with the syndrome compression technique, where $t'=\lceil t_1/(t_1-t_2)\rceil$. Compared with the direct application of the syndrome compression technique, the proposed constructions achieve substantially improved computational efficiency.
	

		\section*{Acknowledgment}
	This work was funded by the European Union through the ERC Advanced
	Grant 101054904: TRANCIDS. Views and opinions expressed are, however,
	those of the authors only and do not necessarily reflect those of the European
	Union or the European Research Council Executive Agency. Neither the
	European Union nor the granting authority can be held responsible for them.
	
	\ifCLASSOPTIONcaptionsoff
	\newpage
	\fi

	\bibliographystyle{IEEEtran}
	\bibliography{myreference}

@string(TIT = "IEEE Transactions on Information Theory")

@string(ISIT = "International Symposium on Information Theory")

@INPROCEEDINGS{liu2026two,
	author={Liu, Yajuan and Duman, Tolga M.},
	booktitle={2026 IEEE International Symposium on Information Theory (ISIT)}, 
	title={Two Bursts of $t_1$-Deletion-$t_2$-Insertion Error-Correcting Codes}, 
	year={2026},
	month = {Jul.}
}

@ARTICLE{jeong2003forward,
	author={Jaein Jeong and Cheng Tien Ee},
	journal={Technical report, Department of Electrical Engineering and Computer Science, University of California, Berkeley}, 
	title={Forward error correction in sensor networks}, 
	year={2003}}

@ARTICLE{venkataramanan2015low,
	author={Venkataramanan, Ramji and Narasimha Swamy, Vasuki and Ramchandran, Kannan},
	journal={IEEE Transactions on Information Theory}, 
	title={Low-Complexity Interactive Algorithms for Synchronization From Deletions, Insertions, and Substitutions}, 
	year={2015},
	volume={61},
	number={10},
	pages={5670-5689},
	month={Oct.}}

@ARTICLE{bours1994construction,
	author={Bours, A.H.},
	journal={IEEE Transactions on Information Theory}, 
	title={Construction of fixed-length insertion/deletion correcting runlength-limited codes}, 
	year={1994},
	volume={40},
	number={6},
	pages={1841-1856},
	month={Nov.}}

@ARTICLE{hu2010achievable,
	author={Hu, Jun and Duman, Tolga M. and Erden, M. Fatih and Kavcic, Aleksandar},
	journal={IEEE Transactions on Communications}, 
	title={Achievable information rates for channels with insertions, deletions, and intersymbol interference with i.i.d. inputs}, 
	year={2010},
	volume={58},
	number={4},
	pages={1102-1111},
	month={Apr.}}

@ARTICLE{paluncic2012multiple,
	author={Paluncic, Filip and Abdel-Ghaffar, Khaled A. S. and Ferreira, Hendrik C. and Clarke, Willem A.},
	journal={IEEE Transactions on Information Theory}, 
	title={A Multiple Insertion/Deletion Correcting Code for Run-Length Limited Sequences}, 
	year={2012},
	volume={58},
	number={3},
	pages={1809-1824},
	month={Mar.}}

@ARTICLE{levenshtein1993perfect,
	author={Levenshtein, V.I. and Vinck, A.J.H.},
	journal={IEEE Transactions on Information Theory}, 
	title={Perfect $(d,k)$-codes capable of correcting single peak-shifts}, 
	year={1993},
	volume={39},
	number={2},
	pages={656-662},
	month={Mar.},
	doi={10.1109/18.212300}}

@book{alon2016probabilistic,
 	title     = {The Probabilistic Method},
 	edition   = {4},
 	author    = {Alon, Noga and Spencer, Joel H.},
 	series    = {Wiley Series in Discrete Mathematics and Optimization},
 	year      = {2016},
 	publisher = {Wiley},
 	location  = {Hoboken, NJ}
 }

@ARTICLE{liu2024explicit,
 	author={Liu, Shu and Tjuawinata, Ivan and Xing, Chaoping},
 	journal={IEEE Transactions on Information Theory}, 
 	title={Explicit Construction of $q$-Ary $2$-Deletion Correcting Codes With Low Redundancy}, 
 	year={2024},
 	volume={70},
 	number={6},
 	pages={4093-4101},
 	month={Jun.},
 	doi={10.1109/TIT.2024.3360964}}

@ARTICLE{song2023nonbinary,
	author={Song, Wentu and Cai, Kui},
	journal={IEEE Transactions on Information Theory}, 
	title={Non-Binary Two-Deletion Correcting Codes and Burst-Deletion Correcting Codes}, 
	year={2023},
	volume={69},
	number={10},
	pages={6470-6484},
	month={Oct.}
}

@ARTICLE{sun2024binary,
	author={Sun, Yubo and Ge, Gennian},
	journal={IEEE Transactions on Information Theory}, 
	title={Binary Codes for Correcting Two Edits}, 
	year={2024},
	volume={70},
	number={10},
	pages={6877-6898},
	month={Oct.},
	doi={10.1109/TIT.2024.3445929}}

@INPROCEEDINGS{lenz2020optimal,
	author={Lenz, Andreas and Polyanskii, Nikita},
	booktitle={2020 IEEE International Symposium on Information Theory (ISIT)}, 
	title={Optimal Codes Correcting a Burst of Deletions of Variable Length}, 
	year={2020},
	pages={757-762},
	month={Jun.},
	doi={10.1109/ISIT44484.2020.9174288}
}

@article{song2022systematic,
	title={Systematic Codes Correcting Multiple-Deletion and Multiple-Substitution Errors},
	author={Wentu Song and Nikita Polyanskii and Kui Cai and Xuan He},
	journal={IEEE Transactions on Information Theory},
	year={2022},
	volume={68},
	number={10},
	pages={6402-6416},
	month={Oct.}
}

@inproceedings{smagloy2020singleseletion,
	title={Single-Deletion Single-Substitution Correcting Codes},
	author={Ilia Smagloy and Lorenz Welter and Antonia Wachter-Zeh and Eitan Yaakobi},
	booktitle={2020 IEEE International Symposium on Information Theory (ISIT)},
	year={2020},
	pages={775-780},
	month={Jun.}
}

@INPROCEEDINGS{lu2022tdeletion1,
	title={$t$-Deletion-$1$-Insertion-Burst Correcting Codes},
	author={Ziyang Lu and Yiwei Zhang},
	booktitle={2022 IEEE International Symposium on Information Theory (ISIT)},
	year={2022},
	pages={808-813},
	month={Jun.}
}

@article{sun2024asymptotically,
	author={Sun, Yubo and Lu, Ziyang and Zhang, Yiwei and Ge, Gennian},
	journal={IEEE Transactions on Information Theory}, 
	title={Asymptotically Optimal Codes for $(t,s)$-Burst Error}, 
	year={2025},
	volume={71},
	number={3},
	pages={1570-1584},
	month={Mar.}
}

@article{ye2024codes,
	title={Codes Correcting Two Bursts of Exactly $b$ Deletions},
	author={Zuo Ye and Yubo Sun and Wenjun Yu and Gennian Ge and Ohad Elishco},
	journal={IEEE Transactions on Information Theory},
	year={2026},
	volume={72},
	number={4},
	pages={2083-2098},
	month={Apr.}
}

@ARTICLE{gabrys2019codes,
	author={Gabrys, Ryan and Sala, Frederic},
	journal={IEEE Transactions on Information Theory}, 
	title={Codes Correcting Two Deletions}, 
	year={2019},
	volume={65},
	number={2},
	pages={965-974},
	month={Feb.},
	doi={10.1109/TIT.2018.2876281}}

@article{sima2020two,
	title={Two Deletion Correcting Codes From Indicator Vectors},
	author={Jin Sima and Netanel Raviv and Jehoshua Bruck},
	journal={IEEE Transactions on Information Theory},
	year={2020},
	volume={66},
	number={4},
	pages={2375-2391},
	month={Apr.}
}

@ARTICLE{sima2021on,
	author={Sima, Jin and Bruck, Jehoshua},
	journal={IEEE Transactions on Information Theory}, 
	title={On Optimal $k$-Deletion Correcting Codes}, 
	year={2021},
	volume={67},
	number={6},
	pages={3360-3375},
	month={Jun.},
	doi={10.1109/TIT.2020.3028702}}

@ARTICLE{schoeny2017codes,
	author={Schoeny, Clayton and Wachter-Zeh, Antonia and Gabrys, Ryan and Yaakobi, Eitan},
	journal={IEEE Transactions on Information Theory}, 
	title={Codes Correcting a Burst of Deletions or Insertions}, 
	year={2017},
	volume={63},
	number={4},
	pages={1971-1985},
	month={Apr.}}

@INPROCEEDINGS{sima2020syndrome,
	author={Sima, Jin and Gabrys, Ryan and Bruck, Jehoshua},
	booktitle={2020 IEEE International Symposium on Information Theory (ISIT)}, 
	title={Syndrome Compression for Optimal Redundancy Codes}, 
	year={2020},
	volume={},
	number={},
	pages={751-756},
	month={Jun.},
	doi={10.1109/ISIT44484.2020.9174009}
}

@ARTICLE{lu2023tdeletion,
	author={Lu, Ziyang and Zhang, Yiwei},
	journal={IEEE Transactions on Information Theory}, 
	title={$t$-Deletion-$s$-Insertion-Burst Correcting Codes}, 
	year={2023},
	volume={69},
	number={10},
	pages={6401-6413},
	month={Oct.},
	doi={10.1109/TIT.2023.3289055}
}

@INPROCEEDINGS{sima2020optimalcodes,
	author={Sima, Jin and Gabrys, Ryan and Bruck, Jehoshua},
	booktitle={2020 IEEE International Symposium on Information Theory (ISIT)}, 
	title={Optimal Codes for the $q$-ary Deletion Channel}, 
	year={2020},
	pages={740-745},
	month={Jun.},
	doi={10.1109/ISIT44484.2020.9174241}
}

@article{guruswami2021explicit,
	title={Explicit Two-Deletion Codes With Redundancy Matching the Existential Bound},
	author={Venkatesan Guruswami and Johan H{\aa}stad},
	journal={IEEE Transactions on Information Theory},
	year={2021},
	volume={67},
	number={10},
	pages={6384-6394},
	month={Oct.}
}

@article{pi2025two,
	author={Pi, Yuhang and Zhang, Zhifang},
	journal={IEEE Transactions on Information Theory}, 
	title={Two-Insertion/Deletion/Substitution Correcting Codes}, 
	year={2025},
	volume={71},
	number={9},
	pages={6743-6758},
	month={Sep.}}

@article{cai2021correcting,
	title={Correcting a Single Indel/Edit for {DNA}-Based Data Storage: Linear-Time Encoders and Order-Optimality},
	author={Kui Cai and Yeow Meng Chee and Ryan Gabrys and Han Mao Kiah and Tuan Thanh Nguyen},
	journal={IEEE Transactions on Information Theory},
	year={2021},
	volume={67},
	number={6},
	pages={3438-3451},
	month={Jun.}
}

@ARTICLE{wang2024nonbinary,
	author={Wang, Shuche and Tang, Yuanyuan and Sima, Jin and Gabrys, Ryan and Farnoud, Farzad},
	journal={IEEE Transactions on Information Theory}, 
	title={Non-Binary Codes for Correcting a Burst of at Most $t$ Deletions}, 
	year={2024},
	volume={70},
	number={2},
	pages={964-979},
	month={Feb.},
	doi={10.1109/TIT.2023.3340246}}

@INPROCEEDINGS{song2024new,
	title={New Construction of $q$-ary Codes Correcting a Burst of at Most $t$ Deletions},
	author={Wentu Song and Kui Cai and Tony Q. S. Quek},
	booktitle={2024 IEEE International Symposium on Information Theory (ISIT)},
	year={2024},
	month={Jul.},
	pages={1101-1106}
}

@ARTICLE{varshamov1965code,
	author={Varshamov, R. R.  and Tenengol'ts, G. M.},
	journal={Avtomat. i Telemekh.}, 
	title={Codes for Correcting a Single Asymmetric Error}, 
	year={1965},
	volume={26},
	number={2},
	pages={288-292}}

@article{schaller2025a,
	title={A New Construction of Non-Binary Deletion Correcting Codes and their Decoding},
	author={Michael Schaller and  Beatrice Toesca and Van Khu Vu},
	journal={arXiv:2501.13534},
	year={2025},
	month={Jan.}
}

@Inproceedings{cheng2019block,
	author =	{Cheng, Kuan and Jin, Zhengzhong and Li, Xin and Wu, Ke},
	title =	{Block Edit Errors with Transpositions: Deterministic Document Exchange Protocols and Almost Optimal Binary Codes},
	volume ={},
	booktitle =	{46th International Colloquium on Automata, Languages, and Programming (ICALP 2019)},
	pages =	{1-15},
	year =	{2019},
	month={Jul.}
}

@ARTICLE{sun2025codes,
	author={Sun, Yubo and Ge, Gennian},
	journal={IEEE Transactions on Information Theory}, 
	title={Codes for Correcting a Burst of Edits Using Weighted-Summation {VT} Sketch}, 
	year={2025},
	volume={71},
	number={3},
	pages={1631-1646},
	month={Mar.}
	}

@article{levenshtein1965binary,
	title={Binary codes capable of correcting deletions, insertions and reversals},
	author={Levenshtein, V. I. },
	journal={Doklady Akademii Nauk SSSR},
	volume = {10},
	number = {8},
	pages = {707-710},
	year={1966},
	month={Feb.}
}

@ARTICLE{brakensiek2018efficient,
	author={Brakensiek, Joshua and Guruswami, Venkatesan and Zbarsky, Samuel},
	journal=TIT,
	title={Efficient Low-Redundancy Codes for Correcting Multiple Deletions},
	year={2018},
	volume={64},
	number={5},
	pages={3403-3410},
	month={May},
	doi={10.1109/TIT.2017.2746566}
}

@article{
	heckel2019acharacterization,
	title={A characterization of the {DNA} data storage channel},
	author={ Heckel, R.  and  Mikutis, G.  and  Grass, Robert N },
	journal={Scientific Reports},
	volume={9},
	pages={1--12},
	year={2019},
	month = {Jul.}
}

@INPROCEEDINGS{sima2020optimal,
author={Sima, Jin and Gabrys, Ryan and Bruck, Jehoshua},
booktitle={2020 IEEE International Symposium on Information Theory (ISIT)}, 
title={Optimal Systematic $t$-Deletion Correcting Codes}, 
year={2020},
volume={},
number={},
pages={769-774},
month={Jun.}}
	
	\begin{IEEEbiography}[{\includegraphics[width=1in,height=1.25in,clip,keepaspectratio]{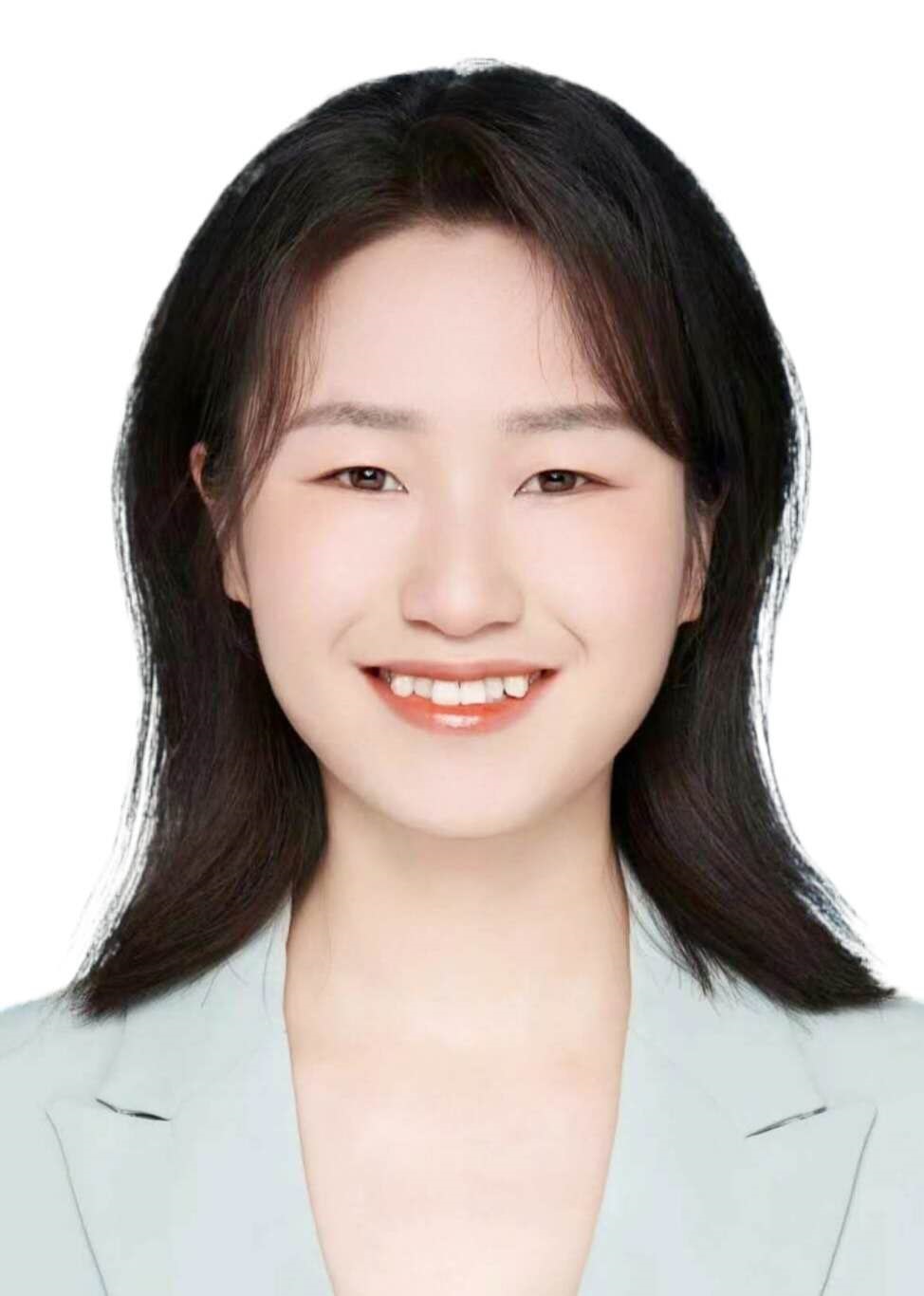}}]{Yajuan Liu} (Member, IEEE) 
		received the B.S. degree in mathematics and applied mathematics from Northwest Normal University, Lanzhou, China, in 2018, and the Ph.D. degree from Southwest Jiaotong University, Chengdu, China, in 2024. She is currently a Postdoctoral Fellow at Bilkent University, Ankara, Turkey. Her research interests include coding for distributed and DNA data storage systems.
	\end{IEEEbiography}
	
	\begin{IEEEbiography}[{{\includegraphics[scale=0.2]{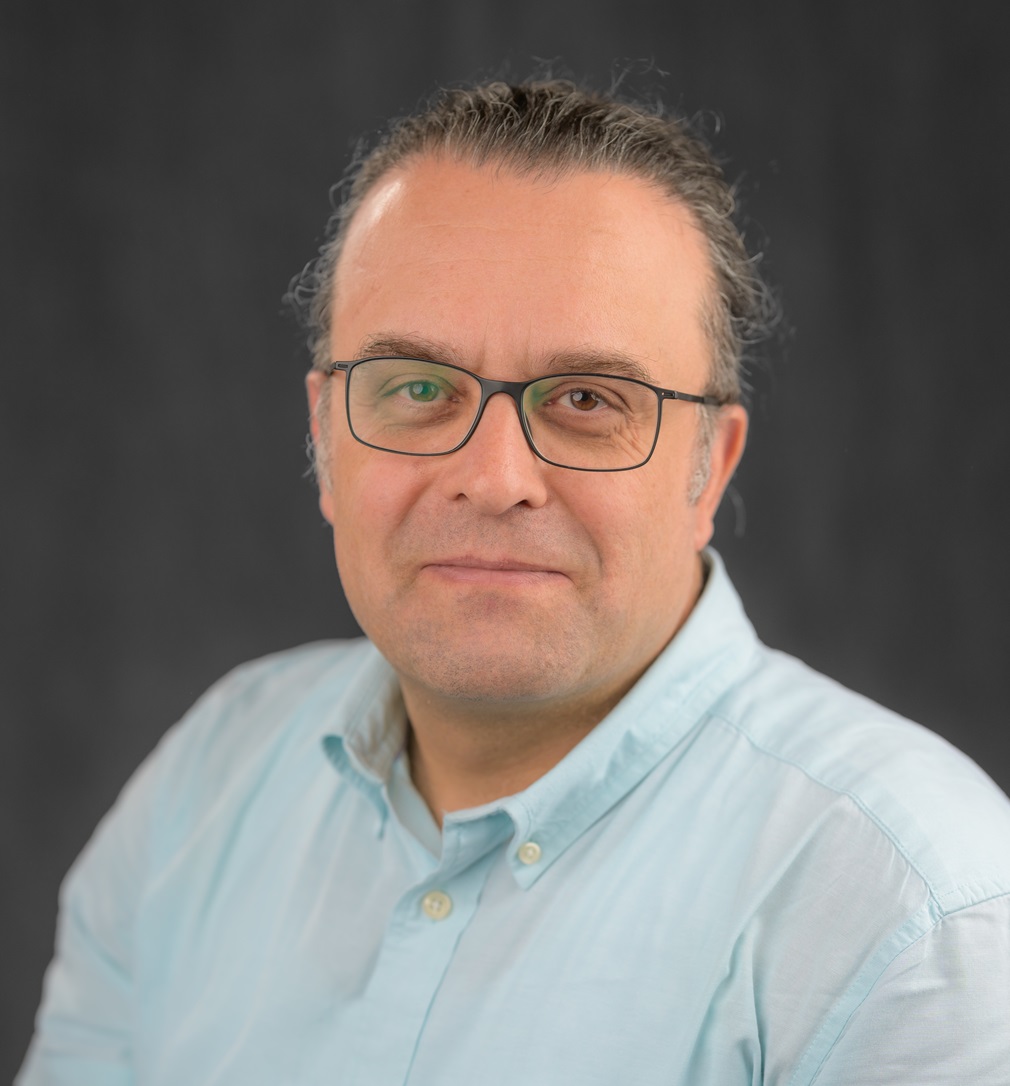}}}]
		{Tolga M. Duman} (S'95--M'98--SM'03--F'11) is a Professor
		in the Electrical and Electronics Engineering Department at Bilkent
		University in Turkey. He received a B.S. degree from Bilkent University,
		Ankara, Turkey, in 1993, and the M.S. and Ph.D. degrees from Northeastern
		University, Boston, MA, USA, in 1995 and 1998, respectively, all in
		electrical engineering. Before joining Bilkent University in September
		2012, he was a full professor with the School of ECEE at Arizona State
		University. Dr. Duman's current research interests are in systems,
		with a particular focus on communications and signal processing, including
		wireless and mobile communications, channel coding/modulation, coding
		for wireless communications, information theory, and data storage systems,
		as well as machine learning for communications and semantic communications/signal
		processing.\\
		Dr. Duman is a Fellow of the IEEE, a recipient of the National Science
		Foundation CAREER Award, the IEEE Third Millennium medal, and a European
		Research Council (ERC) Advanced Grant. He served on the editorial
		boards of various journals, including IEEE TRANSACTIONS ON WIRELESS
		COMMUNICATIONS AND IEEE COMMUNICATIONS SURVEYS AND TUTORIALS. He also
		served as Editor-in-Chief of Elsevier's Physical Communication journal
		and IEEE TRANSACTIONS ON COMMUNICATIONS. 
	\end{IEEEbiography}
\end{document}